\documentclass[lettersize,journal]{IEEEtran}
\usepackage{amsmath,amsfonts}
\usepackage{algorithmic}
\usepackage{algorithm}
\usepackage{array}
\usepackage[caption=false,font=normalsize,labelfont=sf,textfont=sf]{subfig}
\usepackage{stfloats}
\usepackage{url}
\usepackage{verbatim}
\usepackage{indentfirst}
\usepackage{amsfonts}
\usepackage{graphicx}
\usepackage{multirow}
\usepackage{amssymb}
\usepackage{amsmath}
\usepackage{diagbox}
\usepackage{bbding}
\usepackage{makecell}
\usepackage{xcolor}
\usepackage{float}  
\usepackage[breaklinks=true,colorlinks,bookmarks=false]{hyperref}
\usepackage{subfiles}
\usepackage{booktabs}
\usepackage{soul}
\usepackage{cite}
\usepackage{textcomp, gensymb}
\usepackage{ulem}

\usepackage[final]{changes} % 提交时可改为 final 关闭标注
\definecolor{reviewer1}{RGB}{228,26,28}   % 鲜红色
\definecolor{reviewer2}{RGB}{55,126,184}  % 深蓝色
\definecolor{reviewer3}{RGB}{77,175,74}   % 草绿色
\definecolor{reviewer4}{RGB}{152,78,163}  % 紫罗兰
\definecolor{reviewer5}{RGB}{237,125,49}  % 橙色
\definecolor{reviewer6}{RGB}{237,125,49}  % 橙色
\definecolor{reviewer7}{RGB}{237,125,49}  % 橙色

\definechangesauthor[name={Reviewer 1}, color=reviewer1]{R1}
\definechangesauthor[name={Reviewer 2}, color=reviewer2]{R2}
\definechangesauthor[name={Reviewer 3}, color=reviewer3]{R3}
\definechangesauthor[name={Reviewer 4}, color=reviewer4]{ME}
\definechangesauthor[name={Reviewer 5}, color=reviewer5]{R1R2}
\definechangesauthor[name={Reviewer 6}, color=reviewer6]{R1R3}
\definechangesauthor[name={Reviewer 7}, color=reviewer7]{R2R3}

\begin{document}

\title{Self-supervised Learning-based Reconstruction of High-resolution 4D Light Fields}

\author{Jianxin~Lei,
        Dongze~Wu,
        Chengcai~Xu,
        Hongcheng~Gu,
        Guangquan~Zhou,~\IEEEmembership{Member,~IEEE},
        Junhui~Hou,~\IEEEmembership{Senior~Member,~IEEE},
        and~Ping~Zhou,~\IEEEmembership{Member,~IEEE}% <-this % stops a space
\thanks{This work was supported in part by the National Natural Science Foundation of China under Grant 62371121, in part by the NSFC Excellent Young Scientists Fund 62422118, in part by the Hong Kong Research Grants Council under Grant 11218121, and in part by the Hong Kong Innovation and Technology Fund under Grant MHP/117/21. \textit{(Corresponding authors: Junhui Hou and Ping Zhou)}}
\thanks{J. Lei, D. Wu, C. Xu, H. Gu, G. Zhou and P. Zhou are with the School of Biological Science $\&$ Medical Engineering, Southeast University, Nanjing, China. E-mails: leijx@seu.edu.cn; 
 capzhou@163.com.}% <-this % stops a space
\thanks{J. Hou is with the Department of Computer Science, City University of Hong Kong, Hong Kong. E-mail: jh.hou@cityu.edu.hk.}% <-this % stops a space
}
% \markboth{Journal of \LaTeX\ Class Files,~Vol.~14, No.~8, August~2015}%
% \markboth{MANUSCRIPT SUBMITTED TO IEEE TRANSACTIONS ON VISUALIZATION AND COMPUTER GRAPHICS}
% {Shell \MakeLowercase{\textit{et al.}}: Bare Demo of IEEEtran.cls for IEEE Journals}

\maketitle

\begin{abstract}
Hand-held light field (LF) cameras often exhibit low spatial resolution due to the inherent trade-off between spatial and angular dimensions. Existing supervised learning-based LF spatial super-resolution (SR) methods, which rely on pre-defined image degradation models, struggle to overcome the domain gap between the training phase---where LFs with natural resolution are used as ground truth---and the inference phase, which aims to reconstruct higher-resolution LFs, especially when applied to real-world data.
To address this challenge, this paper introduces a novel self-supervised learning-based method for LF spatial SR, which can produce higher spatial resolution LF images than originally captured ones without pre-defined image degradation models. The self-supervised method incorporates a hybrid LF imaging prototype, a real-world hybrid LF dataset, and a self-supervised LF spatial SR framework. The prototype makes reference image pairs between low-resolution central-view sub-aperture images and high-resolution (HR) images. The self-supervised framework consists of a well-designed LF spatial SR network with hybrid input, a central-view synthesis network with an HR-aware loss that enables side-view sub-aperture images to learn high-frequency information from the only HR central view reference image, and a backward degradation network with an epipolar-plane image gradient loss to preserve LF parallax structures.
Extensive experiments on both simulated and real-world datasets demonstrate the significant superiority of our approach over state-of-the-art ones in reconstructing higher spatial resolution LF images without pre-defined degradation. The code and dataset will be publicly available at \url{https://github.com/LeiJianxin/SSLB-HLFSSR}.
\end{abstract}


% \fi
\IEEEpeerreviewmaketitle

%%=============================================================%%
%%======================  Introduction   ======================%%
%%=============================================================%%

\section{Introduction}
\label{sec:intro}
\begin{figure*}[htbp]
    \centering
    \includegraphics[width=0.9\linewidth]{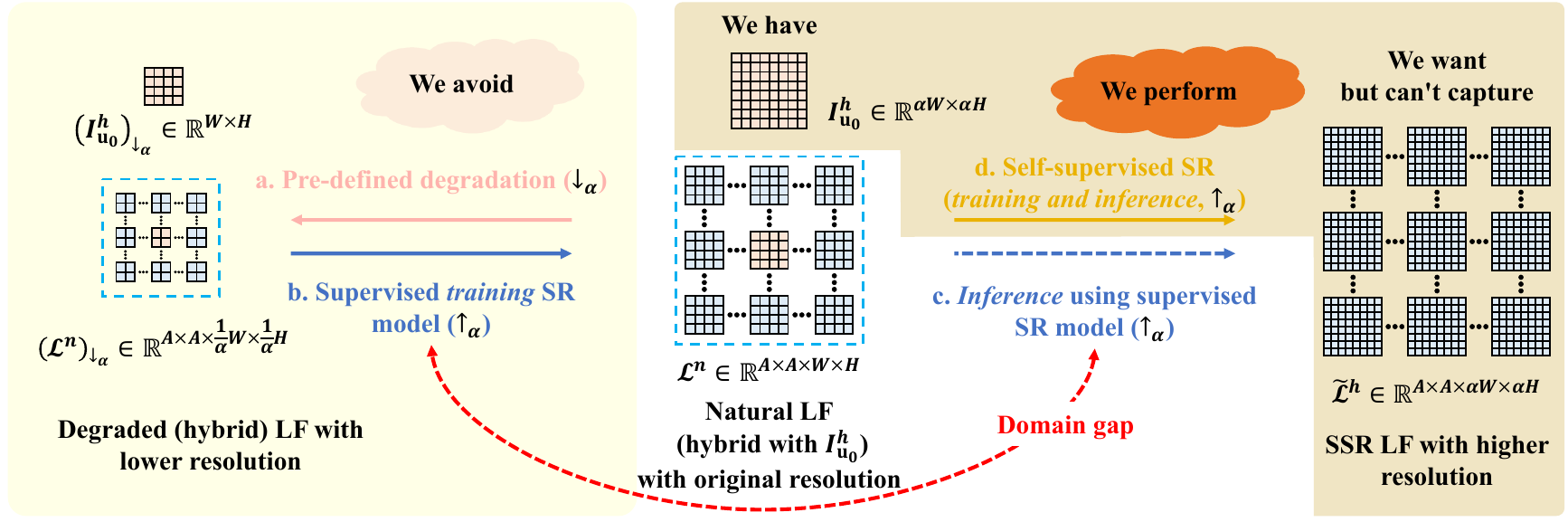}
    \caption{Schematic illustrating the key differences between our proposed method and existing methods in the training and inference phases of LF spatial SR networks. \textbf{Existing methods}: First, training data pairs are created through phase \textit{a}, followed by training the network through phase \textit{b}, and finally applying the network trained in phase \textit{b} for inference in phase \textit{c}. \textbf{Our method} (phase \textit{d}): Without the need for pre-defined degradation and HR LF ground truth, the outcomes of network training and inference are both LF images in the target HR domain.}
    \label{fig:motivation}
\end{figure*}

\IEEEPARstart{L}{ight} Field (LF) cameras, which can simultaneously record the density and direction of light rays, have found extensive applications in areas such as digital refocusing \cite{yang2023Hybridlenses,zhao2024refocusing}, 3-dimensional (3D) reconstruction \cite{cai2018ray, Zhou20223d,jin2022occlusion,zhou2023light}, and virtual reality \cite{yu2017vr}. However, due to the limited resolution of image sensors in commercial LF cameras and the trade-off between spatial and angular resolution, the further development of LF cameras is constrained. Acquiring higher spatial resolution LF images requires more expensive and precise camera components. Therefore, the LF spatial super-resolution (SR) algorithms are both essential and economically valuable for the broader adoption of LF cameras.
\par
Significant advancements in LF spatial SR methods have been achieved in recent years. These methods are generally categorized into non-learning-based methods \cite{Bishop2012,Wanner2014,Farrugia2017,Alain2018,Rossi2018,boominathan2014PaSR,wang2016iPADs}, supervised learning-based methods \cite{yoon2015LFSR,yuan2018LFSREPI,wang2018LFNet,zhang2019resLF,yeung2018LFSSR,jin2020AlltoOne,Guo2022DRLF,wang2022disentangling,chen2023LFSSR-HI,jin2020LFHSR,chang2022FHLFSR,jin2023LFHSR,wang2024real-worldLF, lyu2024probabilistic}, and self-supervised learning-based methods \cite{cheng2021LFZSSR,sheng2023lfassr}. Most existing supervised and self-supervised methods are developed based on the assumption that the LF image degradation model is known and fixed. As illustrated in phase \textit{a} in Figure \ref{fig:motivation}, these methods begin by artificially degrading the LF images in natural resolution to generate low-resolution (LR) and ``high-resolution (HR)"\footnote{This ``HR" is not the actual target resolution, but the natural resolution.} LF image training pairs, utilizing either a single fixed degradation (e.g., bicubic down-sampling) or multiple assumed degradations (e.g., as proposed by Wang \textit{et al.} \cite{wang2024real-worldLF}). In the following, we collectively refer to these as pre-defined image degradation. These supervised SR networks are then trained using the LR-``HR" LF image pairs, as illustrated in phase \textit{b} in Figure \ref{fig:motivation}. While these methods continually innovate and evolve at the network architecture, achieving near-saturated SR performance under pre-defined degradation scenarios, they inherently face a critical limitation: pre-defined degradations cannot accurately simulate unknown degradations, especially in real-world LF spatial SR tasks\footnote{Real-world SR tasks aim to produce higher-resolution images than the natural, often without higher-resolution ground truth.}. When actual degradation deviates from these pre-defined degradations---such as when the trained SR network (phase \textit{b}) is applied to generate higher spatial resolution (phase \textit{c}), especially on the real images with unknown and complex degradations---a significant domain gap emerges.
For example, the spatial resolution of the sub-aperture images (SAIs) recorded by the current Lytro ILLUM \cite{Lytro} cameras is about $620\times 430$, which serves as the target resolution for most SR methods. These methods typically down-sample the LF SAIs by $2\times$ or $4\times$, and aim to super-resolve them back to the target resolution of $620\times 430$. However, when these SR networks are employed to obtain higher-resolution SAIs, such as $1240\times 860$ ($2\times$SR) or $2480\times1920$ ($4\times$SR), a domain gap inevitably arises. Moreover, as there is the usual lack of LF images with the resolution of $1240\times 860$ or $2480\times1920$ as ground truth, these methods cannot train networks directly from original to target resolution, and instead, they must rely on the network's generalizability.
\par
In this paper, our objective is to eliminate the dependence on predefined degradation priors and learn the SR process by implicitly modeling the practical degradation of the LF image, as shown in the phase \textit{d} in Figure \ref{fig:motivation}. In the single image super-resolution (SISR) community, a commonly used strategy to achieve implicit modeling of the image degradation process is unsupervised learning GAN-based SR methods\cite{yuan2018CinCGAN,Wei2021DASR,liu2023BlindSR}.
However, due to the lack of HR ground truth for supervision, the quantitative performance of these unsupervised methods is usually inferior to that of supervised methods in SR tasks with pre-defined degradation. 
In real-world LF spatial SR tasks without paired HR ground truth, the challenge of implicitly modeling the LF image degradation lies in how to appropriately provide adequate information that describes the degradation process during the training phase, which differs from the training of supervised methods providing corresponding HR images for each LR SAI. Since the LF SAIs can be seen as images captured from different views of one scene, one of the SAIs may contain most degradation information for the LF spatial SR tasks. Thus, if we can capture an LR-HR SAI pair from one specific view (such as the central view), then for the other LR SAIs, a non-supervised SR network can be designed to achieve LF spatial SR by leveraging the central view's LR SAI-HR image reference pairs without estimating the degradation model.
\par
It should be noted that, on the one hand, the proposed LF spatial SR backbone is similar to the existing LF spatial SR networks with hybrid input \cite{jin2023LFHSR, chang2022FHLFSR}, thus our method is classified as a hybrid LF spatial SR method. However, it differs from the existing hybrid LF spatial SR methods \cite{jin2023LFHSR, chang2022FHLFSR} that rely on pre-defined degradation and perform the phases \textit{a-c} in Figure \ref{fig:motivation} in the absence of actual HR ground truth with target resolution. Our method implicitly models the LF image degradation process without paired HR LF image training data and achieves high-quality SR reconstruction of real-world LF images with unknown and complex degradation, as shown in phase $d$ in Figure \ref{fig:motivation}. On the other hand, we categorize our method in this paper as a self-supervised method, as the global input and output of our LF spatial SR training framework remain consistent. To the best of our knowledge, there are no methods except ours that do not rely on pre-defined degradation assumptions and are based on non-supervised learning for LF spatial SR. Furthermore, in theory, once a target HR image of a certain view is provided, our self-supervised framework can be used to train the LF spatial SR network for enhancing LR LF images, with the learned image degradation reversal process being closer to actual conditions.
\par
In this paper, to achieve higher spatial resolution than that of the original LF dataset without relying on pre-defined degradation and paired ground truth, we have carried out work from three dimensions: hardware system, dataset, and deep learning-based LF spatial SR network. The main contributions of this paper are as follows:
\begin{enumerate}
    \item We developed and calibrated a hybrid LF imaging prototype. As shown in Figure \ref{fig:system}, this prototype mainly consists of a micro-lens array (MLA)-based LF camera, an HR 2D digital single-lens reflex (DSLR) camera, and a beam splitter, enabling simultaneous capture of 4D LF images and HR 2D images. The central SAI of the LF and the HR 2D image form the aforementioned central view LR SAI-HR image reference pairs. 
    \item We built a real-world hybrid LF dataset using our prototype. This dataset contains 189 scenes of 4D LR LF and their corresponding 2D HR reference images. The real image degradation model implicit in the HR reference image pairs provides important degradation information for the following self-supervised framework.
    \item We proposed a novel self-supervised LF spatial SR framework that includes an LF spatial SR network with hybrid input (HLFSSR-Net), a central view synthesis network (CVS-Net), and a backward degradation network (BD-Net) to achieve the goal of LF spatial SR without pre-defined degradation and paired ground truth.
\end{enumerate}
\par
Experiments on both simulated LF datasets and our real-world LF dataset have validated the effectiveness of our method. A common degradation model \cite{wang2022disentangling,chang2022FHLFSR,jin2023LFHSR} is used to form the simulated datasets, while the degradation model of our dataset is unknown and complex. 
\par
The rest of this paper is organized as follows. Section \ref{sec:related_work} reviews related works on LF spatial SR methods. Section \ref{sec:pre} analyzes the LF spatial SR issue with an implicit degradation model in theory and introduces our solution strategy. Section \ref{sec:system} presents the proposed hybrid LF imaging prototype and the real-world hybrid LF dataset captured by this prototype. Section \ref{sec:net} describes our self-supervised LF spatial SR framework in detail. Sections \ref{sec:experiments} and \ref{sec:exp_real} present extensive experiments on both simulated LF datasets and our real-world dataset with unknown degradation to evaluate our method. Finally, Section \ref{sec:conclusion} concludes this paper.

%%=============================================================%%
%%======================  Related Work   ======================%%
%%=============================================================%%
\section{Related Work}
\label{sec:related_work}
\subsection{Hybrid LF Imaging}
Hybrid LF imaging is proposed to address issues such as the narrow baseline and the trade-off between spatial and angular resolution in hand-held MLA-based LF cameras. Based on the optical path relationship between 4D LF cameras and 2D DSLR cameras, hybrid LF imaging systems are mainly classified into the parallel optical axis systems \cite{boominathan2014PaSR,wang2017lfvideoHybrid,alam2018hybrid,ye2023LFIENet}, and the shared optical axis systems \cite{wang2016hrHybrid, xiong2017hybrid}. The parallel optical axis system has a simple structure, but it is difficult to align the field of view (FOV) of LF cameras and DSLR cameras. The shared optical axis system overcomes the viewpoint drawback via the beam-splitter, aligning the FOV of the 4D LF camera with that of the 2D DSLR camera. In addition, Wang \textit{et al.} \cite{wang2016iPADs} presented a hybrid LF imaging system via an array-based LF imaging scheme, which is equipped with a central HR 2D camera surrounded by eight LR side-view USB cameras. The array-based hybrid LF imaging system overcomes both the short baseline and the trade-off of LF resolution, and achieves notable results in LF spatial SR \cite{wang2016iPADs,jin2020LFHSR,chang2022FHLFSR,jin2023LFHSR}. However, this approach still presents inherent drawbacks. For example, the whole system is bulky, and the accurate system calibration and refinement are complicated. 

\subsection{LF Spatial SR}
\subsubsection{Without Hybrid Input} LF spatial SR aims to improve the spatial resolution of all SAIs. Early non-learning-based methods \cite{Bishop2012,Wanner2014,Farrugia2017,Alain2018,Rossi2018} produced suboptimal results with blur and distortion. Recently, learning-based methods for LF spatial SR have made significant progress \cite{yoon2015LFSR,yuan2018LFSREPI,wang2018LFNet,zhang2019resLF,yeung2018LFSSR,jin2020AlltoOne, Guo2022DRLF, wang2022disentangling}.
Yoon \textit{et al.} \cite{yoon2015LFSR} proposed the first CNN-based algorithm (LFCNN) to learn the correspondence among stacked SAIs for LF images SR, but LFCNN neglected the angular relationship of LF. Wang \textit{et al.} \cite{wang2022disentangling} organized SAIs into macro-pixels and proposed a disentangling mechanism-based LF spatial SR network (named DistgSSR) that fully explored the 4D information of LF. 
Recently, Liang \textit{et al.} \cite{liang2022LFT} applied Transformers to LF image SR, and then proposed EPIT \cite{liang2023EPIT} to learn the non-local spatial-angular correlation for LF image SR. 
Although these supervised methods have achieved increasingly superior SR performance through increasingly advanced network architectures under the assumption of using bicubic or bilinear interpolation as image degradation models, their application in introducing higher-resolution real-world LF remains challenging. This is due to the lack of sufficient HR ground truth and the domain gap between real-world test data (with realistic and complex degradation models) and simulated training data (with assumed and simple degradation models).
\par
\subsubsection{With Hybrid Input} With the rise of the hybrid LF imaging system, LF reconstruction with a hybrid input has become a promising way. Boominathan \textit{et al.} \cite{boominathan2014PaSR} proposed a patch matching-based algorithm (named PaSR) to improve the LF spatial resolution through hybrid input. Wang \textit{et al.} \cite{wang2016iPADs} improved the resolution of side-view SAIs by iterative refinement combining patch-based SR results with depth-based synthesis (named iPADS). Zheng \textit{et al.} \cite{zheng2017hybrid} combined an exemplar-based approach and a learning-based approach for LF SR. Zhao \textit{et al.} \cite{zhao2018HCSR} designed a reference-based method that transferred the details from the HR image to the LR side-view SAIs to obtain high-quality LF. Recently, Chang \textit{et al.} \cite{chang2022FHLFSR} used layered refinement to design an LF SR network with hybrid input, demonstrating clear advantages in cases of large disparity ranges. Jin \textit{et al.} \cite{jin2020LFHSR,jin2023LFHSR} proposed a framework with two complementary and parallel research lines, namely SR-Net and Warp-Net, and their advantages were combined via attention-guided fusion.  Additionally, Jin \textit{et al.} \cite{jin2023LFHSR} used color perturbation to augment the training dataset, addressing color inconsistency in the real-world hybrid dataset \cite{wang2016iPADs}.
\par
For LF SR with hybrid input, the SR results obtained by the non-learning-based methods \cite{boominathan2014PaSR,wang2016iPADs,zheng2017hybrid} still suffer from blur and distortion. In contrast, the supervised methods \cite{chen2023LFSSR-HI,jin2020LFHSR,chang2022FHLFSR,jin2023LFHSR} often achieve acceptable SR performance. However, the supervised methods have to rely on sufficient HR LF ground truth and pre-defined image degradation. Therefore, the challenge for LF SR remains in achieving higher spatial resolution in real-world conditions without pre-defined degradation.

\subsection{Real-world LF Spatial SR}
Self-supervised learning (SSL) aims to construct surrogate supervisory signals directly from the data itself, without relying on manually annotated labels \cite{Gui2024SSL, Chen_2024_CVPR}. To reduce reliance on paired HR LF ground truth and on pre-defined degradation models, recent studies have advanced zero-shot and self-supervised strategies for LF SR. For instance, Cheng \textit{et al.} \cite{cheng2021LFZSSR} introduced a zero-shot LF SR framework that trains directly on the input LF to mitigate the training--inference domain gap. Sheng \textit{et al.} \cite{sheng2023lfassr} proposed a cross-view recurrence-based self-supervised LF SR model that eliminates the need for HR supervision by leveraging angular recurrence among SAIs to guide spatial detail restoration.
\par
Nevertheless, many of these methods continue to assume simplified or parameterized degradation models, which limits their effectiveness when confronted with the complex, spatially varying degradations produced by real optical systems. Recently, research has focused on improving generalization capability to real-world degradation. Xiao \textit{et al.} \cite{xiao2023real-worldLF} created the LytroZoom dataset containing paired real-world LR-HR LF images and proposed a novel baseline network named OFPNet for real-world LF SR work. The SR performance of OFPNet in LytroZoom is superior to the existing methods, but the spatial resolutions of HR ground truth in the LytroZoom dataset are limited to $456\times320$ and $608\times416$. To enhance generalization capability to real-world degradation, Wang \textit{et al.} \cite{wang2024real-worldLF} formulated a more practical LF degradation model based on bicubic interpolation with isotropic Gaussian blur and additive noise of multiple kernel widths and noise levels. They also proposed a degradation-modulating network (LF-DMnet), which exhibits acceptable generalization. However, the multiple degradations simulated by the LF-DMnet still cannot fully represent more complex real-world cases.

%%=============================================================%%
%%===================    preliminaries      ===================%%
%%=============================================================%%
\begin{figure*}[htbp]
    \centering
    \includegraphics[width=0.85\linewidth]{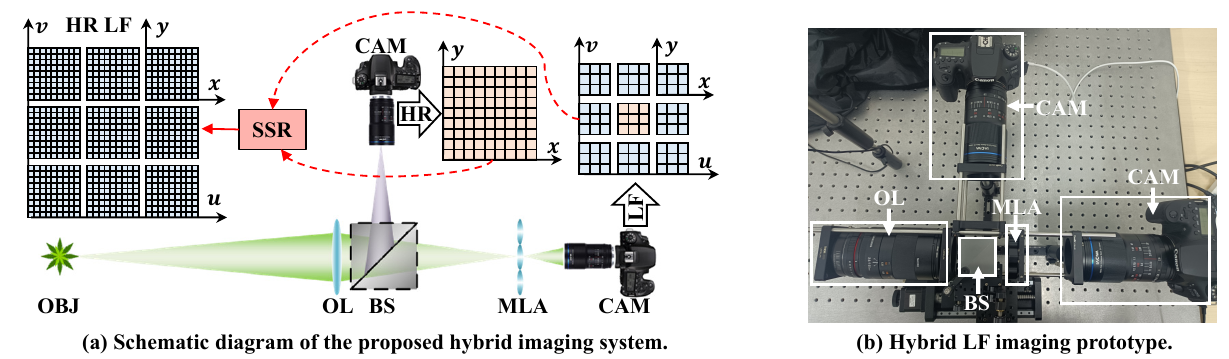}
    \vspace{-1.0em}
    \caption{Illustration of our hybrid 4D LF imaging prototype. OBJ: object, OL: objective lens, BS: beam splitter, MLA: micro-lens array, CAM: camera, HR: 2D high spatial resolution image, LF: 4D LF image with low spatial resolution, SSR: LF spatial SR with hybrid input, HR LF: 4D LF image with higher spatial resolution. }
    \label{fig:system}
\end{figure*}

\section{Preliminaries}
\label{sec:pre}
In this paper, we use the two-plane model \cite{Levoy1996LFRendering} to parameterize 4D natural LF as $\mathcal{L}^n(\mathbf{u}, \mathbf{x})\in \mathbb{R}^{M \times N \times W \times H} $ with size of $M \times N \times W \times H$, where $\mathbf{u} = \{(u,v) \mid 1 \leq u \leq M, 1 \leq v \leq N \}$ represents angular coordinates, $\mathbf{x}=\{(x,y) \mid 1 \leq x \leq W, 1 \leq y \leq H \}$ represents spatial coordinates, as shown in Figure \ref{fig:motivation}. Let $I^n_{\mathbf{u}_0} \in \mathbb{R}^{ W \times H}$ represents the central SAI with spatial resolution of $W \times H$ and $I_{\mathbf{u}_0}^h \in \mathbb{R}^{\alpha W \times \alpha H}$ denotes its corresponding 2D HR reference image, where $\mathbf{u}_0=(u_0, v_0)$ is the angular coordinate of the central SAI, and $\alpha$ is an up-sample scale factor. It should be noted that most LF SR methods use SAIs distributed in a square array as their inputs, so we set $M = N = A$ in this paper, where $A\times A$ denotes the angular resolution. 

\subsection{LF Image Degradation Model}
In optical imaging systems, image degradation is typically caused by diffraction, aberrations, and specific camera structures. These degradation factors are usually modeled by a point spread function (PSF) that causes image blurring, the sampling density of the sensor, and some additional noise. For LF cameras based on MLA, SAIs are viewed as images of the scene from different viewpoints $\mathbf{u} = (u, v)$, and their degradation model is expressed by Equation \ref{eq:degradation}. Typically, SR is considered to be the inverse process of solving Equation \ref{eq:degradation}.
\begin{equation}
    {I}_\mathbf{u}^{n}=\left({I}_\mathbf{u}^{h} \otimes K_\mathbf{u}\right) \downarrow_\alpha+{N}_\mathbf{u},
    \label{eq:degradation}
\end{equation}
where $\otimes$ denotes convolution operation, ${I}_{\mathbf{u}}^{n} \in \mathbb{R}^{W \times H} $ denotes LR SAI in 4D LF $\mathcal{L}$ with angular coordinate $\mathbf{u} = (u, v)$ and ${I}_{\mathbf{u}}^{h}  \in \mathbb{R}^{\alpha W \times \alpha H} $ denotes its corresponding ideal HR SAI, $\alpha$ is defined as the up-sampling or down-sampling factor. This degradation model has three key components: the blur kernel matrix $K_{\mathbf{u}}$, the down-sampling operation $\downarrow_\alpha$, and the additive noise matrix $N_{\mathbf{u}}$. 
\par
According to the taxonomy by Liu \textit{et al.} \cite{liu2023BlindSR}, most learning-based methods in the LF spatial SR community belong to non-blind SR ones. These methods use the image degradation model based on Equation \ref{eq:degradation} to artificially degrade existing LF images $\mathcal{L}^n$ to generate lower resolution LF images, as shown in phase \textit{a} in Figure \ref{fig:motivation}. Most methods \cite{wang2022disentangling,chang2022FHLFSR,jin2023LFHSR} consider only the $\downarrow_\alpha$ factor in their degradation model, while a few \cite{wang2024real-worldLF} also consider the three factors: $K_{\mathbf{u}}$, $\downarrow_\alpha$ and $N_{\mathbf{u}}$. 
\par
\subsection{Problem Formulation}
As shown in Figure \ref{fig:motivation}, there are LF images $\mathcal{L}^n \in \mathbb{R}^{ A \times A \times  W \times  H}$ with a spatial resolution of $ H \times W $. In hybrid LF imaging systems, there is also 2D HR image $I_{\mathbf{u}_0}^h \in \mathbb{R}^{\alpha W \times  \alpha H} $, where the spatial resolution of $\alpha H \times \alpha W$ is the target of SR task. In the LF spatial SR task, the goal is to reconstruct the LF images $\tilde{\mathcal{L}}^h$ with a higher-resolution of $\alpha H \times \alpha W$, which is $\alpha$ times higher than that of $\mathcal{L}^n$. Since there is no corresponding ground truth for $\tilde{\mathcal{L}}^h$, the supervised methods first apply pre-defined degradation to down-sample the LF images $\mathcal{L}^n$ by a factor of $\alpha$ according to Equation \ref{eq:degradation}. Consequently, the supervised methods obtain $\left(\mathcal{L}^n\right)_{\downarrow_\alpha} \in \mathbb{R}^{ A \times A \times  \frac{1}{\alpha} W \times \frac{1}{\alpha} H}$ and $\left(I_{\mathbf{u}_0}^h\right)_{\downarrow_\alpha} \in \mathbb{R}^{ W \times H} $ as input, as shown in phase \textit{a} in Figure \ref{fig:motivation}. Here, the degraded 2D HR image $\left(I_{\mathbf{u}_0}^h\right)_{\downarrow_\alpha}$ serves as an auxiliary reference during supervised training, enabling the network to learn the mapping between the LR LF input and the high-frequency details implicitly encoded in the central view. This design follows prior hybrid LF SR approaches~\cite{jin2023LFHSR,chang2022FHLFSR}. Subsequently, these supervised $\alpha \times$ SR networks are trained end-to-end, 
% the SR networks are designed and trained, 
as shown in phase \textit{b} in Figure \ref{fig:motivation}, expressed as,
\begin{equation}
    \tilde{\mathcal{L}}^n=G_{\overline{SR}}\left((\mathcal{L}^n)_{\downarrow_\alpha},(I_{\mathbf{u}_0}^h)_{\downarrow_\alpha}\right),
    \label{eq:trainESLB}
\end{equation}
where the subscript $\overline{SR}$ indicates that the network is based on supervised learning. Typically, these methods solve the SR network $G_{\overline{SR}}$ with the following optimization goal:
\begin{equation}
    \begin{aligned}
    &arg \min\left|\tilde{\mathcal{L}}^n-\mathcal{L}^n\right|\\
    =&arg \min\left|G_{\overline{SR}}\left((\mathcal{L}^n)_{\downarrow_\alpha},(I_{\mathbf{u}_0}^h)_{\downarrow_\alpha}\right)- \mathcal{L}^n\right|.
    \end{aligned}
    \label{eq:optimizeESLB}
\end{equation}
\par
Finally, the SR network $G_{\overline{SR}}$ is used to produce $\tilde{\mathcal{L}}^h$, as shown in phase \textit{c} in Figure \ref{fig:motivation}, represented as:
\begin{equation}
    \tilde{\mathcal{L}}^h=G_{\overline{SR}}\left(\mathcal{L}^n,I_{\mathbf{u}_0}^h\right).
    \label{eq:inferenceESLB}
\end{equation}
\par
As shown in phases \textit{a} to \textit{c} in Figure \ref{fig:motivation} and Equations \ref{eq:trainESLB} and \ref{eq:inferenceESLB}, although the $G_{\overline{SR}}$ is used for the \textbf{$\alpha \times$} SR task from $\mathcal{L}^n$ to $\tilde{\mathcal{L}}^h$, it essentially learns the inverse degradation model from $(\mathcal{L}^n)_{\downarrow_\alpha}$ to $\mathcal{L}^n$. Therefore, a domain gap exists between the inverse degradation models in phases $b$ and $c$ in Figure \ref{fig:motivation}. If the inverse degradation models in training and inference phases are similar, the supervised methods could produce acceptable SR results $\tilde{\mathcal{L}}^h$; otherwise, they cannot.
\par
In contrast, our self-supervised approach eliminates the need for pre-defined degradations and one-to-one HR LF ground truth. Both training and inference (Equation \ref{eq:HLFSSR}) directly operate on the non-degraded $\mathcal{L}^n$ and the HR central view $I_{\mathbf{u}_0}^h$, leveraging the latter as an internal high-frequency prior to guide the reconstruction of all SAIs.
To this end, we introduce a self-supervised LF spatial SR framework that directly learns to obtain HR $\tilde{\mathcal{L}}^h$ from hybrid inputs of $\mathcal{L}^n$ and $I_{\mathbf{u}_0}^h$, as shown in phase $d$ in Figure \ref{fig:motivation}, where the LF image degradation is modeled implicitly. The spatial SR network in our self-supervised framework is expressed as: 
\begin{equation}
    \tilde{\mathcal{L}}^h=G_{SR}\left(\mathcal{L}^n,I_{\mathbf{u}_0}^h\right),
    \label{eq:HLFSSR}
\end{equation}
where the SR network $G_{SR}$ without the subscript of $\overline{SR}$ indicates that our method is not an supervised one. Compared with the supervised methods, our approach does not encounter a domain gap issue, as the inverse degradation from $\mathcal{L}^n$ to $\tilde{\mathcal{L}}^h$ is modeled implicitly and learned directly. Without one-to-one ground truth for LF SAIs, it is essential for different view SAIs to learn their inverse degradation models from the single 2D HR image $I_{\mathbf{u}_0}^h$ within our SR framework. Therefore, we design the CVS-Net ($G_{CVS}$), the BD-Net ($G_{BD}$), and the HLFSSR-Net ($G_{SR}$) based on the characteristics of the hybrid LF hardware system and dataset in this paper. Furthermore, the HLFSSR-Net can be addressed by solving the following optimization problem, expressed as,
\begin{equation}
\begin{aligned}
&\arg\min_{\hat{I}_{\mathbf{u}_0}^h,\;\hat{\mathcal{L}}^n}\Bigl(
\bigl|\hat{I}_{\mathbf{u}_0}^h - I_{\mathbf{u}_0}^h\bigr|
\;+\;
\bigl|\hat{\mathcal{L}}^n - \mathcal{L}^n\bigr|
\Bigr) \\
&= \arg\min_{G_{SR}}\Bigl(
\bigl|\,G_{CVS}^*\bigl(G_{SR}(\mathcal{L}^n,\,I_{\mathbf{u}_0}^h)\;-\;\tilde{I}_{\mathbf{u}_0}^h\bigr)
\;-\;I_{\mathbf{u}_0}^h\bigr| \\
&\qquad\quad
+\;\bigl|\,G_{BD}^*\bigl(G_{SR}(\mathcal{L}^n,\,I_{\mathbf{u}_0}^h)\bigr)
\;-\;\mathcal{L}^n\bigr|
\Bigr),
\end{aligned}
\label{eq:optimizeOurs}
\end{equation}
where * indicates the frozen parameters. More details about the CVS-Net, BD-Net, HLFSSR-Net and the Equation \ref{eq:optimizeOurs} will be described in Section \ref{sec:net}.
%%=============================================================%%
%%======================     System      ======================%%
%%=============================================================%%
\section{Hybrid 4D LF Imaging Prototype}
\label{sec:system}
\subsection{Setup}
The proposed hybrid LF imaging prototype is mainly equipped with an objective lens, a cubic beam splitter, a co-located low spatial resolution LF imaging subsystem, and a high spatial resolution traditional imaging subsystem, as shown in Figure \ref{fig:system}(a). The light emitted by the object point is received and concentrated by the objective lens. It is then divided into two parts (transmitted light and reflected light) by the beam-splitting plane of the cubic beam splitter. The transmitted part is received by a micro-lens array (MLA) and projected onto the DSLR camera to record the 4D LF image with low spatial resolution 2D SAIs. Simultaneously, the reflected part is projected onto another DSLR camera to record the high spatial resolution 2D traditional image.
\par
As shown in Figure \ref{fig:system}(b), the prototype consists of the objective lens (focal length 100mm, F2.8), the cubic beam splitter (cube size 50mm, reflection/transmission ratio 30R/70T), an MLA (focal length 2mm, microlens pitch 0.063mm), two relay lenses (focal length 100mm, F2.8) and two photosensors (Canon 90D). All these modules are mounted via kinematic mounts (Thorlabs) to allow fine adjustment with respect to the optical axis. The size of the 2D HR image is $6960\times4640$ pixels, and the size of the microlens image is $19\times19$ pixels, so the size of the 2D SAI is $329\times217$ pixels. In addition, the optical structure of this prototype is sufficiently simple for integration, offering potential hardware support to overcome the resolution trade-off in hand-held LF cameras.
\par
\subsection{Calibration}
We accomplish the calibration of the hybrid imaging system via the checkerboard, whose pattern is $5\times8$ and grid size is $18\times 18$ mm. First, the LF imaging subsystem is calibrated by Zhou's method \cite{zhou2019calibration}, and the traditional imaging subsystem is calibrated by Zhang's method \cite{zhangzhengyou} respectively. Then, the parameters related to the beam splitter are calibrated by the relationship between the external parameters of the two subsystems. After calibration, the mean re-projection error of the LF and 2D traditional imaging system is $0.139$ and $1.78$ pixels, respectively. The optical axis of the 2D traditional imaging system intersects the main-lens plane of the LF system via the beam splitter, and the angle between the optical axis and the normal of the main-lens plane is $0.501\degree$. The calibration results show that the hybrid imaging system is capable of capturing pairs of 4D LR LF images and 2D HR images simultaneously.
\par
\begin{figure}[!t]
    \centering
    \includegraphics[width=\linewidth]{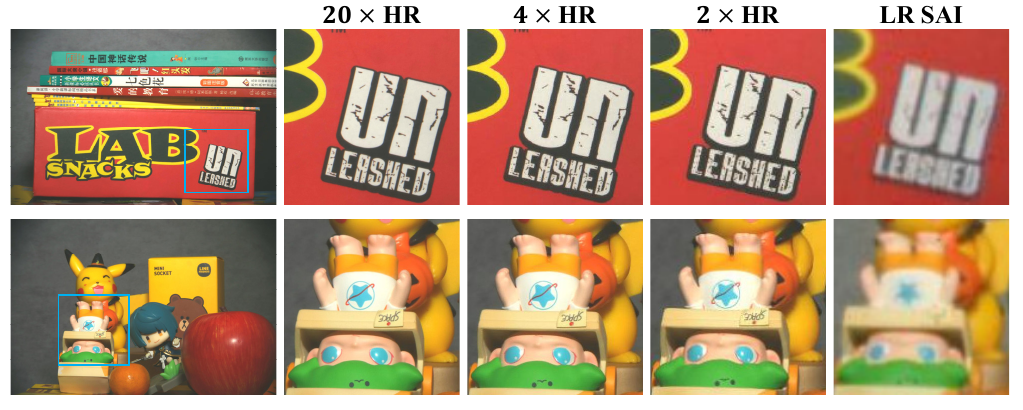}
    % \vspace{-2.0em}
    \caption{Examples of LR LF and different resolution 2D HR image pairs from the proposed hybrid LF dataset. The resolution of LR SAI is $329\times217$ and the resolution of the corresponding 2D HR reference images can be set as from $658\times434$ to $6580\times4340$, which can be used for $2\times$ to $20\times$ SR. }
    \label{fig:dataset}
\end{figure}
\subsection{Dataset Acquisition and Pre-processing}
We acquire a real-world hybrid LF dataset consisting of paired 4D LR LF images and 2D HR images. The dataset contains 189 pairs hybrid images, among which the 2D HR image is the corresponding HR reference version of the central SAI of LF. To maintain the view consistency between the 2D HR image and the central SAI of LF, we compute the affine transformation matrix via feature points in two corresponding images, and we crop and scale the 2D HR image to make the reference image of the central SAI. 
Our hybrid LF dataset includes samples such as colorful books, boxes, blocks, flowers, fruits, and toys. To ensure the color consistency between 2D HR images and LF images, we use the LSA method \cite{ashraf2018cc} to calculate the color correction matrix by capturing the color chart model (Datacolor SpyderCheckr 24). Figures \ref{fig:dataset} and \ref{fig:compareOur} show some samples of our dataset. Compared with LR SAI, the 2D HR reference images have richer detailed information. 

%%=============================================================%%
%%======================     Network     ======================%%
%%=============================================================%%
\section{Proposed Method}
\label{sec:net}

\subsection{Overview}
\begin{figure*}[htbp]
    \centering
    \includegraphics[width=\linewidth]{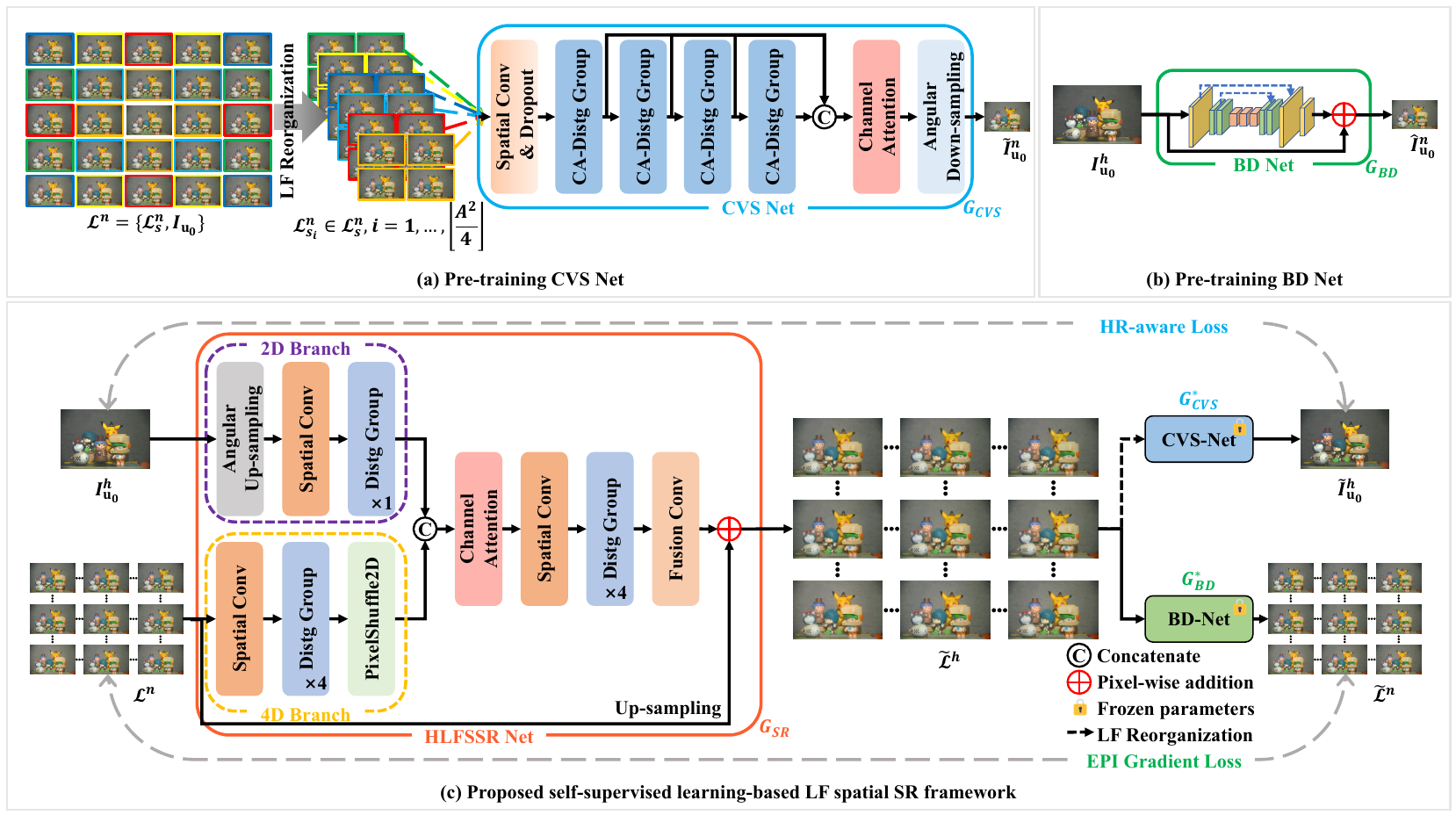}
    \vspace{-2.0em}
    \caption{Illustration of our self-supervised LF spatial SR framework. A $5\times5$ LF is used as an example for illustration. 
    }
    \label{fig:framework}
\end{figure*}

We used the hybrid LF imaging prototype proposed in Section \ref{sec:system} to capture 4D LR LF images $\mathcal{L}^n$ and 2D HR images $I_{\mathbf{u}_0}^h$, where the 2D HR image is the corresponding HR reference version of the central SAI in 4D LR LF. $\mathcal{L}^n$ and $I_{\mathbf{u}_0}^h$ are simultaneously fed into our spatial SR network to produce the spatial super-resolved LF $\tilde{\mathcal{L}}^h$, as shown in Figure \ref{fig:framework}. The entire self-supervised LF spatial SR framework mainly consists of three sub-networks: the LF central view synthesis network (CVS-Net), the backward degradation network (BD-Net), and the LF spatial SR network with hybrid input (HLFSSR-Net).
\par
As shown in Figure \ref{fig:framework}(c), the HLFSSR-Net expressed by Equation \ref{eq:HLFSSR} accomplishes the feature extraction from both of $\mathcal{L}^n$ and $I_{\mathbf{u}_0}^h$, learns the residual map from LR LF $\mathcal{L}^n$ to the spatial super-resolved LF $\tilde{\mathcal{L}}^h$, and then adds the up-sampled results of $\mathcal{L}^n$ to the residual map to obtain $\tilde{\mathcal{L}}^h$. In comparison with supervised methods, our self-supervised framework needs to propagate high-frequency information from only one HR reference image to all side-view SAIs and preserve LF parallax structures due to the lack of one-to-one ground truth for each SAI of $\tilde{\mathcal{L}}^h$.
According to the characteristics of our hybrid LF dataset, the sole HR image $I_{\mathbf{u}_0}^h$ corresponding to the LR central SAI serves as the only HR reference image available for updating the HLFSSR-Net's parameters.
Therefore, we have to construct a supervised pathway with the side-view SAIs as input and expect to gain the central SAI as output. As shown in Figure \ref{fig:framework}(a), we design the CVS-Net that synthesizes the side-view SAIs into the central SAI. In this paper, the CVS-Net is pre-trained and used after the HLFSSR-Net, where its parameters are frozen, as shown in Figure \ref{fig:framework}(c). Then, the first loss function is built between the input of HLFSSR-Net ($I_{\mathbf{u}_0}^h$) and the output of CVS-Net ($\tilde{I}_{\mathbf{u}_0}^h$), which is named as HR-aware Loss, as shown in \ref{fig:framework}(c). 
In addition, in the LF spatial SR task, it is essential to maintain the LF structure, but only the HR-aware Loss is inadequate for this purpose. As the epipolar-plane image (EPI) is a hybrid expression of LF in both the angular and spatial coordinates, we introduce another loss function named EPI Gradient Loss \cite{jin2020AlltoOne,jin2020LFASR}, as shown in Figure \ref{fig:framework}(c). To construct EPI from $\tilde{\mathcal{L}}^h$, whose dimensions differ from $\mathcal{L}^n$, we propose the BD-Net to degrade the SR LF $\tilde{\mathcal{L}}^h$ back to LR LF ${\mathcal{L}}^n$. 
More details about the three sub-networks and the two loss functions are described in the following sub-sections. And the detailed architectures in the networks are shown in supplementary file.
\subsection{LF Central View Synthesis Network}
As mentioned above, the CVS-Net and the HR-aware loss are used to assist the HLFSSR-Net in learning the high-frequency information from 2D HR reference image $I_{\mathbf{u}_0}^h$, where the LF reorganization module and channel-wise attention-based disentangling mechanism are introduced, as shown in Figure \ref{fig:framework}(a).
\par
\subsubsection{LF Reorganization} As shown in Figure \ref{fig:framework}(a), more than one side-view SAI is fed into the CVS-Net with only one central view SAI obtained. According to the characteristics of LF imaging, the spatial information contained in all side-view SAIs far exceeds that in the central SAI for the purpose of central view synthesis. Therefore, the CVS-Net may tend to focus on a small part of the side-view SAIs closer to the central SAI but neglect the contributions from the others. It is necessary to mitigate LF information redundancy for the CVS-Net as it leads to uneven back-propagation from the central SAI to side-view SAIs. Therefore, the LF Reorganization module is introduced in this paper.
\par
Without loss of generality, we suppose that the CVS-Net receives $A^2-1$ side-view SAIs as inputs, where $A$ is the angular resolution of 4D LR LF. As shown in Figure \ref{fig:framework}(a), we reorganize all side-view SAIs $\mathcal{L}^n_s$ into $\lfloor \frac{A \times A}{2\times2} \rfloor$ groups of $2\times2$ SAIs as the input of the CVS-Net, i.e., $\mathcal{L}^n_s = \{\mathcal{L}^n_{s_k}, k = 1, ..., \lfloor \frac{A^2}{4} \rfloor\}$. In the pre-training stage, $\{\mathcal{L}^n_{s_k}\}$ are randomly selected and fed into the CVS-Net, and the CVS-Net is pre-trained to learn the mapping relationship between the side-view SAIs $\mathcal{L}^n_s$ and the central SAI $\tilde{I}^n_{\mathbf{u}_0}$, i.e.,
\begin{equation}
    \tilde{I}^n_{\mathbf{u}_0}=G_{CVS}\left(\mathcal{L}^n_s\right)=G_{CVS}\left(\mathcal{L}^n_{s_1}, ..., \mathcal{L}^n_{s_k}\right), k = \lfloor \frac{A^2}{4} \rfloor.
    \label{eq:CVS}
\end{equation}
\par
The loss function of the CVS-Net is designed to minimize the absolute error between $\tilde{I}^n_{\mathbf{u}_0}$ and the central SAI ${I}^n_{\mathbf{u}_0}$, which is expressed as follow:
\begin{equation}
    \ell_{CVS}=\frac{1}{H \times W} \sum_{\mathbf{x}}\left|I^n_{\mathbf{u}_0}(\mathbf{x})-\tilde{I}^n_{\mathbf{u}_0}(\mathbf{x})\right|,   
    \label{eq:cvsloss}
\end{equation}
where $\mathbf{x}=(x, y), 1 \leq x \leq  W, 1 \leq y \leq H$. So far, we have obtained the pre-trained CVS-Net ($G_{CVS}$).
\par 
\subsubsection{Network Design} We designed the CVS-Net based on the LF disentangling mechanism \cite{wang2022disentangling} and channel-wise attention \cite{hu2018CAM}. As shown in \ref{fig:framework}(a), the CVS-Net mainly consists of four channel-wise attention-based disentangling groups (i.e., CA-Distg Groups), and the output features of each group are concatenated, then fed into the channel-wise attention \cite{hu2018CAM} module for multi-stage feature fusion. Moreover, to improve the generalization ability of the CVS-Net, we introduce Dropout \cite{srivastava2014dropout} (with the dropout probability set to 0.5) at the beginning convolutional layer of the CVS-Net. At the end of the CVS-Net, we apply the $A \times A$ convolutional layer (with the stride set to $A$) and two $1 \times 1$ convolutional layers for angular down-sampling to obtain the central SAI. The detailed architectures in CVS-Net are shown in the supplementary file. 
\par
\subsection{Backward Degradation Network}
As shown in Figure \ref{fig:framework}(c), the BD-Net and the EPI gradient loss are used to guide the HLFSSR-Net in preserving the LF parallax structure corresponding to the input LF $\mathcal{L}^n$. According to the Lambertian model, the light ray emitted by a target scene is recorded by different SAIs of a 4D LF from various perspectives, and the light ray’s intensity should be consistent across the SAIs \cite{lyu2024probabilistic}. And since LF SAIs are captured in a certain scene, the expected high-frequency information in SR results of different views is isotropic. Therefore, the degradation process of the central view learned by the pre-trained BD-Net can provide an important reference to side views. As shown in Figure \ref{fig:framework}(b), we design the BD-Net by establishing the mapping from 2D HR image $I_{\mathbf{u}_0}^h$ to the central SAI $\hat{I}^n_{\mathbf{u}_0}$ based on the U-Net structure in CBD-Net \cite{guo2019CBDNet}, i.e.,
\begin{equation}
    \hat{I}^n_{\mathbf{u}_0}=G_{BD}\left(I_{\mathbf{u}_0}^h\right).
    \label{eq:BD}
\end{equation}
\par
The BD-Net is pre-trained by minimizing the absolute error between $\hat{I}^n_{\mathbf{u}_0}$ and the central SAI ${I}^n_{\mathbf{u}_0}$:
\begin{equation}
    \ell_{BD}=\frac{1}{H \times W} \sum_{\mathbf{x}}\left|I^n_{\mathbf{u}_0}(\mathbf{x})-\hat{I}^n_{\mathbf{u}_0}(\mathbf{x})\right|,
    \label{eq:bdloss}
\end{equation}
where $\mathbf{x}=(x, y), 1 \leq x \leq  W, 1 \leq y \leq H$. We have thus obtained the pre-trained BD-Net ($G_{BD}$).
\par
\subsection{Hybrid LF Spatial SR Network}
We design the HLFSSR-Net ($G_{SR}$) based on the characteristics of the hybrid input. As shown in Figure \ref{fig:framework}(c), the HLFSSR-Net mainly consists of two branches: the 2D branch processes the 2D HR image $I_{\mathbf{u}_0}^h$, and the 4D branch handles the 4D LR LF image $\mathcal{L}^n$. The detailed architectures of the modules in HLFSSR-Net are shown in the supplementary file. 
\par
% \noindent\textbf{2D Branch}. 
\subsubsection{2D Branch} The spatial resolution of the 2D HR image $I_{\mathbf{u}_0}^h$ is significantly higher than that of the 4D LR LF image $\mathcal{L}^n$. This implies that $I_{\mathbf{u}_0}^h$ contains high-frequency information absent in $\mathcal{L}^n$, crucial for enhancing the resolution of the LR side-view SAIs without ground truth. As shown in Figure \ref{fig:framework}(c), we utilize a $3\times 3$ convolutional layer to extract intermediate HR spatial features with $A\times A$ channels, and we reshape the features for angular up-sampling. Subsequently, we use a Spatial-Conv module and a residual disentangling group (Distg Group) \cite{wang2022disentangling} to learn the deep representation of the high-frequency detail information.
\par
\subsubsection{4D Branch} In the 4D branch, four cascaded Distg Groups are mainly used to extract the structure features of the LF $\mathcal{L}^n$. Subsequently, we apply the sub-pixel convolutional layer \cite{shi2016pixelShuffle} to up-sample the extracted features to the target spatial resolution, matching the resolution of the HR LF $\tilde{\mathcal{L}}^h$. 
\par
\subsubsection{Hybrid LF Features Extraction and Fusion} After extracting features in 2D and 4D branches, we combine them via concatenation and feed them into the channel-wise attention module for initial fusion. Then, we utilize a sequence of a Spatial-Conv module, four Distg Groups, and a Fusion-Conv module \footnote{Spatial-Conv and Fusion-Conv modules are detailed in supplementary file.} to learn the high-frequency residual map. Finally, we add the residual map to the up-sampled LR LF images by the bicubic interpolation to obtain the HR LF images $\tilde{\mathcal{L}}^h$. The problem of reconstructing $\tilde{\mathcal{L}}^h$ from the hybrid input via our HLFSSR-Net can be implicitly formulated as Equation \ref{eq:HLFSSR}.
\par
\subsection{Loss Functions}
The optimization problem shown in Equation \ref{eq:optimizeOurs} is used to drive the training of the HLFSSR-Net ($G_{SR}$), leveraging the pre-trained $G_{CVS}$ and $G_{BD}$ as bridges.
\par
To learn high-frequency details of the scene from the 2D HR image $I_{\mathbf{u}_0}^h$, the side-view SAIs in the super-resolved LF $\tilde{\mathcal{L}}^h$ are reorganized and then fed into the pre-trained CVS-Net (with frozen parameters) to generate the central SAI $\hat{I}_{\mathbf{u}_0}^h$, i.e.,
\begin{equation}
    \hat{I}_{\mathbf{u}_0}^h=G_{CVS}^*\left(\tilde{\mathcal{L}_s^h}\right)=G_{CVS}^*\left( G_{SR}\left(\mathcal{L}^n, I_{\mathbf{u}_0}^h\right) - \tilde{I}_{\mathbf{u}_0}^h \right).
    \label{eq:SR_CVS}
\end{equation}
\par
As shown in Figure \ref{fig:framework}(c), we propose the HR-aware loss, which computes the absolute error between $I_{\mathbf{u}_0}^h$ and $\hat{I}_{\mathbf{u}_0}^h$, as one of the loss function for training the HLFSSR-Net:
\begin{equation}
    \ell_{hr}=\frac{1}{\alpha H \times \alpha W} \sum_{\mathbf{x}}\left|I_{\mathbf{u}_0}^h(\mathbf{x})-\hat{I}_{\mathbf{u}_0}^h(\mathbf{x})\right|,
    \label{eq:HRLoss}
\end{equation}
where $\mathbf{x}=(x, y), 1 \leq x \leq \alpha W, 1 \leq y \leq \alpha H$.
\par
Moreover, to maintain the LF parallax structure in the super-resolved LF $\tilde{\mathcal{L}}^h$, $\tilde{\mathcal{L}}^h$ is fed into the pre-trained BD-Net (with frozen parameters) to produce $\hat{\mathcal{L}}^n$, whose spatial resolution is identical to that of $\mathcal{L}^n$, i.e.,
\begin{equation}
    \hat{\mathcal{L}}^n=G_{BD}^*\left(\tilde{\mathcal{L}^h}\right) = G_{BD}^*\left(G_{SR}\left(\mathcal{L}^n, I_{\mathbf{u}_0}^h\right)\right).
    \label{eq:SR_BD}
\end{equation}
\par
Subsequently, we introduce the EPI gradient loss \cite{jin2020AlltoOne,jin2020LFASR}, which computes the $\ell_{1}$ distance between the gradient of EPIs of $\hat{\mathcal{L}}^n$ and $\mathcal{L}^n$, for the training of the HLFSSR-Net. The gradients are computed along both spatial and angular dimensions on both horizontal and vertical EPIs:
\begin{equation}
    \begin{array}{r}
\ell_{epi}=\frac{1}{(N-1) \times \alpha H}\sum_{y, v}\left(\left|\nabla_x E_{y, v}(x, u)-\nabla_x \hat{E}_{y, v}(x, u)\right|\right. \\
\left.+\left|\nabla_u E_{y, v}(x, u)-\nabla_u \hat{E}_{y, v}(x, u)\right|\right) \\
+\frac{1}{(M-1) \times \alpha W}\sum_{x, u}\left(\left|\nabla_y E_{x, u}(y, v)-\nabla_y \hat{E}_{x, u}(y, v)\right|\right. \\
\left.+\left|\nabla_v E_{x, u}(y, v)-\nabla_v \hat{E}_{x, u}(y, v)\right|\right) ,
\end{array}
    \label{eq:EPILoss}
\end{equation}
where $\nabla$ is the differential operator, $E_{y, v}=\mathcal{L}^n_s\left(x, y^*, u, v^*\right)$, $E_{x, u}=\mathcal{L}^n_s\left(x^*, y, u^*, v\right)$, $\hat{E}_{y, v}=\hat{\mathcal{L}}^n_s\left(x, y^*, u, v^*\right)$ and $\hat{E}_{x, u}=\hat{\mathcal{L}}^n_s\left(x^*, y, u^*, v\right)$ denote the vertical and horizontal EPIs of the side-view SAIs $\mathcal{L}^n_s$ and $\hat{\mathcal{L}}^n_s$.
\par
The aforementioned two loss functions, the HR-aware loss in Equation \ref{eq:HRLoss} and the EPI gradient loss in Equation \ref{eq:EPILoss}, jointly guide the HLFSSR-Net to update its parameters in the absence of the HR ground truth of the side-view SAIs. The HLFSSR-Net is trained by minimizing the combined loss of equal weighting \cite{jin2020AlltoOne, jin2023LFHSR, jin2022LFASR}:
\begin{equation}
    \ell = \ell_{hr} + \ell_{epi}
    \label{eq:totalLoss}
\end{equation}
%%=============================================================%%
%%======================   Experiments   ======================%%
%%=============================================================%%
\section{Experiments on Simulated Datasets}
\label{sec:experiments}
\par
\subsection{Experimental Setup}
\label{subsec:exp_setup}
Since no published papers employ methods similar to ours, a completely fair comparison using standard experimental setups from existing literature is challenging. Therefore, we adopted the following experimental conditions: first, we take the spatial resolution of the original LF SAIs from the public datasets as the target resolution for spatial SR. In other words, we treat the original LF images as the SSR LF images $\mathcal{L}^h$, which is something we want but can't capture, 
% the LF with the original resolution acts as the SSR LF $\mathcal{L}^h$ we want but can't capture, 
as shown in Figure \ref{fig:motivation}. Meanwhile, we mark the central SAI of $\mathcal{L}^h$ as ${I}_{\mathbf{u}_0}^h$. Then $\mathcal{L}^h$ is down-sampled by a factor of $\alpha$ using bicubic interpolation to produce $\mathcal{L}^n$. Consequently, the simulated datasets are constructed that consist of 4D LR LF images $\mathcal{L}^n$ and 2D HR images ${I}_{\mathbf{u}_0}^h$ acts as the natural (hybrid) LF we have, as shown in Figure \ref{fig:motivation}. It should be noted that although there are side-view SAIs with target spatial resolution, they cannot be used as ground truth for any methods, but only be used to calculate quantitative metrics. This setup simulates real-world conditions where no HR LF ground truth is available, facilitating comparisons with various state-of-the-art (SOTA) methods in addressing simple unknown degradation models.
\par
We use both synthetic LF datasets (HCI new \cite{honauer2017HCI} and DLFD \cite{shi2019DLFD} datasets) and real-world LF datasets (EPFL \cite{rerabek2016EPFL} and INRIA \cite{le2018INRIA} datasets captured by Lytro ILLUM camera \cite{Lytro}, and the STFgantry captured by Lego Gantry \cite{vaish2008STFgantry}) to make the simulated datasets for experiments. We partitioned the HCI new and DLFD datasets for training and testing according to Jin \textit{et al.} \cite{jin2023LFHSR}, and followed Wang \textit{et al.} \cite{wang2022disentangling} for EPFL, INRIA, and STFgantry. We follow \cite{yeung2018LFSSR,wang2022disentangling,jin2023LFHSR} to convert the RGB LF images to YCbCr space and only use the Y components for training and quantitative evaluation. The Cb and Cr components are up-sampled by bicubic interpolation for the visual evaluation. We conduct the above experimental setup on all LFs in these datasets, whose angular resolution is $9 \times 9$. For simplicity, we only present the results on $5 \times 5$ LFs for $2 \times$ and $4 \times$ SR. We employ a variety of SOTA methods from the LF spatial SR community for a comprehensive evaluation of our approach: 
\begin{itemize}
    \item \textit{PaSR} \cite{boominathan2014PaSR} and \textit{iPADS} \cite{wang2016iPADs}: Two SOTA non-learning-based LF spatial SR methods with hybrid input.
    \item  \textit{LF-DMnet} \cite{wang2024real-worldLF}: The SOTA supervised real-world LF spatial SR method relies on image degradation prior.
    \item \textit{FHLFSR} \cite{chang2022FHLFSR} and \textit{LFHSR} \cite{jin2023LFHSR}: Two SOTA supervised LF image SR methods with hybrid input. 
    \item \textit{LFZSSR} \cite{cheng2021LFZSSR}: The only self-supervised method with training code released, but without hybrid input.
\end{itemize}
\par

\begin{table*}[htbp]
\centering
\renewcommand\arraystretch{1.25}
    \caption{Quantitative comparisons of different SR methods in terms of the number of parameters (\#Param.) and PSNR($\uparrow$)/SSIM($\uparrow$)/LPIPS($\downarrow$) for $2 \times$ SR on simulated hybrid data. The best results are marked with \uline{underlines}. The difference in average scores between our method and the other methods is placed behind the corresponding average scores. The superscripts $\dag$ and $\S$ represent different input modes: hybrid LF and LF only.
    }
    \vspace{-0.3cm}
    \label{tab:exp_comparison}
    \resizebox{\textwidth}{!}
    {
    \centering
    \begin{tabular}{c|c|c|c|c|c|c|c|c}
    \toprule[1.2pt]
    \multicolumn{2}{c|}{Methods}

    & \#Param.
    & HCI new	
    & DLFD	
    & EPFL	
    & INRIA	
    & STFgantry 
    & Average 
    \\
    
    \hline
    \multirowcell{3}{Non-\\Learning}
    & Bicubic\textsuperscript{\S}
    & -
    & 30.86/0.929/0.148					
    & 33.65/0.954/0.120
    & 29.50/0.935/0.142	
    & 31.10/0.956/0.130
    & 30.82/0.947/0.118 	
    & 31.19({-7.18})/0.944({-0.040})/0.132(+0.126) 
    \\	 
    ~
    & PaSR\textsuperscript{\dag} \cite{boominathan2014PaSR}	
    & -
    & 32.27/0.935/0.151	
    & 35.44/0.963/0.112 
    & 30.22/0.929/0.164 
    & 32.44/0.956/0.141
    & 31.47/0.956/0.114
    & 32.37({-6.00})/0.948({-0.036})/0.136(+0.130)
    \\			
    ~
    & iPADS\textsuperscript{\dag} \cite{wang2016iPADs}	
    & -
    & 36.68/0.978/0.014					
    & 38.99/0.985/0.010
    & 33.69/0.971/0.017
    & 34.69/0.973/0.018
    & 36.54/0.988/0.012
    & 36.12({-2.25})/0.979({-0.005})/0.014(+0.008)
    \\
   
    \hline
    \multirow{3}*{Supervised} 
    
    & LF-DMnet\textsuperscript{\S} \cite{wang2024real-worldLF}	
    & 3.75M						
    & 34.22/0.963/0.061		
    & 38.58/0.982/0.057
    & 32.31/0.965/0.049	
    & 33.04/0.976/0.047
    & 32.84/0.970/0.049 
    & 34.20({-4.17})/0.971({-0.013})/0.053(+0.047) 
    \\	
     ~
    & FHLFSR\textsuperscript{\dag} \cite{chang2022FHLFSR}	
    & 0.54M
    & 39.91/0.989/0.007	
    & 41.77/0.991/0.007
    & 35.08/0.980/0.019
    & 36.94/0.984/0.023
    & 38.87/0.994/0.005
    & {38.51}({+0.14})/\uline{0.988}({+0.004})/0.012(+0.006) 
    \\	
    ~
    & LFHSR\textsuperscript{\dag} \cite{jin2023LFHSR}	
    & 10.79M
    & 40.28/0.988/0.008
    & 42.21/0.989/0.009
    & 35.26/0.983/0.017	
    & 37.12/0.986/0.022
    & 39.02/0.993/0.007
    & \uline{38.78}({+0.41})/\uline{0.988}({+0.004})/0.013(+0.007) 	
    \\	
    
    \hline
    \multirowcell{2}{Self-\\Supervised}
    & LFZSSR\textsuperscript{\S} \cite{cheng2021LFZSSR}	
    & 4.89M
    & 32.34/0.948/0.079					
    & 35.15/0.964/0.055
    & 31.62/0.956/0.060
    & 33.39/0.970/0.048
    & 32.54/0.964/0.044
    & 33.01({-5.36})/0.960({-0.024})/0.057(+0.051)
    \\	
    ~
    & \textbf{Ours}\textsuperscript{\dag}	
    & 7.97M
    & \textbf{39.29}/\textbf{0.984}/\textbf{0.005}			 
    & \textbf{41.30}/\textbf{0.987}/\textbf{0.005}	
    & \textbf{35.23}/\textbf{0.976}/\textbf{0.008}
    & \textbf{36.68}/\textbf{0.978}/\textbf{0.009}
    & \textbf{39.36}/\textbf{0.993}/\textbf{0.004}
    & \textbf{38.37}/\textbf{0.984}/\uline{\textbf{0.006}}				
    \\

    \bottomrule[1.2pt]
    
    \end{tabular}
    }
\end{table*}
\par

\begin{table}[htbp]
\centering
\renewcommand\arraystretch{1.25}
    \caption{Quantitative comparisons of FHLFSR, LFHSR and Ours in terms of PSNR($\uparrow$)/SSIM($\uparrow$)/LPIPS($\downarrow$)/the average SSIM($\uparrow$) of EPIs for $4 \times$ SR on simulated hybrid data. 
    }
    \vspace{-0.3cm}
    \label{tab:exp_comparison3}
    \resizebox{\linewidth}{!}
    {
    \centering
    \begin{tabular}{c|c|c|c}    
    \toprule[1.2pt]
    
    Methods
    & FHLFSR\textsuperscript{\dag} \cite{chang2022FHLFSR}
    & LFHSR\textsuperscript{\dag} \cite{jin2023LFHSR}
    & \textbf{Ours}\textsuperscript{\dag}	

    \\    
    \hline
    
    \#Param.
    & 0.66M
    & 10.98M
    & 7.98M
    
    \\
    \hline
    HCI new	
    & 33.26/0.968/0.018/0.966	
    & 31.84/0.945/0.026/0.944
    & \textbf{35.30}/\textbf{0.969}/\textbf{0.009}/\textbf{0.968}
		
    \\	    
    DLFD	
    & 32.10/0.956/0.059/0.957
    & 30.67/0.922/0.051/0.930  
    & \textbf{36.20}/\textbf{0.966}/\textbf{0.008}/\textbf{0.970}
      			
    \\				
    EPFL	
    & 33.37/0.971/0.017/0.967				 
    & 31.72/0.964/0.030/0.957	
    & \textbf{33.57}/\textbf{0.968}/\textbf{0.012}/\textbf{0.965}

    \\    
    INRIA	
    & 29.27/0.958/0.043/0.943
    & 33.79/0.972/0.035/0.966
    & \textbf{35.55}/\textbf{0.972}/\textbf{0.014}/\textbf{0.970}

    \\    
    STFgantry 
    & 33.22/0.963/0.054/0.958
    & 26.83/0.918/0.046/0.921
    & \textbf{34.60}/\textbf{0.977}/\textbf{0.011}/\textbf{0.976}

    \\ 
    \hline
    \multirow{2}*{Average}
     
    & 32.24/0.963/0.038/0.958 	
    & 30.97/0.944/0.038/0.944	
    & \multirow{2}*{\uline{\textbf{35.04}}/\uline{\textbf{0.970}}/\uline{\textbf{0.011}}/\uline{\textbf{0.970}}}
    \\
    
    ~
    & ({-2.80}/{-0.007}/{+0.027}/{-0.012})	
    & ({-4.07}/{-0.026}/{+0.027}/{-0.026})	
    & ~	
    \\
    \bottomrule[1.2pt]
    
    \end{tabular}
    }
\end{table}
The task for all methods, including ours, is to produce unknown SSR LF $\mathcal{L}^h$ from the natural LF $\mathcal{L}^n$ (and ${I}_{\mathbf{u}_0}^h$ when a hybrid dataset is used), as shown in Figure \ref{fig:motivation}. 
For supervised methods, we perform the phase $a$ to down-sample the natural LF $\mathcal{L}^n$ and ${I}_{\mathbf{u}_0}^h$ by a factor of $\alpha$ using bicubic interpolation. In phase $b$, the natural LF $\mathcal{L}^n$ serves as the ground truth for end-to-end training of these supervised learning-based $\alpha \times$ SR networks. Finally, the trained networks execute phase $c$ to produce SSR LF $\mathcal{L}^h$. For fair comparisons, we used the strategies described in their respective papers to retrain the supervised networks, and the retrained models achieved comparable performance to those provided by the authors.
\subsection{Implementation Details}
\label{subsec:implement}
In our training framework, we crop each SAI into patches of $32\times32$ pixels, crop each corresponding 2D HR reference image into patches of $64\times64$ pixels (for $2\times$ SR) and $128\times128$ pixels (for $4\times$ SR).
We perform random horizontal and vertical flipping, and 90-degree rotation in spatial and angular dimensions to augment the training data. 
We adopt Adam optimizer \cite{kingma2014adam} with $\beta_{1} = 0.9$, $\beta_{2} = 0.999$ for both the pre-training and training. The learning rate of all networks is set to $2\times10^{-4}$ initially and decreases by a factor of 0.5 every 15 epochs. The training of the BD-Net is stopped after 100 epochs. The training of both the CVS-Net and the HLFSSR-Net is stopped after 50 epochs. Our networks are implemented in PyTorch on a PC with an NVIDIA RTX 3090 GPU.
\par
\subsection{Comparisons with State-of-the-art Methods}
\label{subsec:compareSOTA}
\subsubsection{Comparisons of Quantitative Results} To comprehensively evaluate the performance of our method, we use PSNR, SSIM, and LPIPS \cite{zhang2018LPIPS} for quantitative comparisons with different SOTA methods. The experimental results for $2\times$ and $4\times$ SR tasks are presented in Tables \ref{tab:exp_comparison} and \ref{tab:exp_comparison3} respectively. 
\par
For the $2\times$ SR task, our method outperforms all other SOTA methods in average LPIPS scores and excels in PSNR and SSIM scores over all non-learning-based and self-supervised method LFZSSR \cite{cheng2021LFZSSR}. Compared with the supervised method LF-DMnet \cite{wang2024real-worldLF} with LF images as input only, our method surpasses it by more than $4$ dB. Compared with the other two supervised methods with hybrid inputs, while FHLFSR \cite{chang2022FHLFSR} and LFHSR \cite{jin2023LFHSR} obtain better PSNR and SSIM results, our method closely trails by mere margins of $0.14$ dB and $0.41$ dB, respectively. In $2\times$ SR task, the reason why FHLFSR and LFHSR exhibit better results than ours may be that the domain gap between phases $b$ and $c$ shown in Figure \ref{fig:motivation} is small, the acceptable $2\times$ SR performance are also achieved when the networks trained in phase $b$ are used in phase $c$. 
In addition, we conducted $4\times$ SR experiments with a larger domain gap between phases $b$ and $c$, to showcase our method's robustness against large domain gaps. There is no doubt that our method achieves better scores in all metrics in comparison with non-learning-based, self-supervised (LFZSSR), and one of the supervised (LF-DMnet) methods, so only the results in comparison with FHLFSR and LFHSR are depicted in Table \ref{tab:exp_comparison3}. Different from $2\times$ SR results, FHLFSR and LFHSR exhibit difficulties in handling large domain gaps. Our method significantly outperforms FHLFSR and LFHSR, and achieves much higher PSNR scores than LFHSR by $4.07$ dB and FHLFSR by $2.80$ dB. This indicates that our self-supervised method is able to accomplish LF spatial SR without pre-defined degradation, i.e., we can handle the large domain gap better than these two supervised methods. We refer readers to the supplementary file for quantitative results on domain gap.
\begin{figure*}[htbp]
    \centering
    \includegraphics[width=\linewidth]{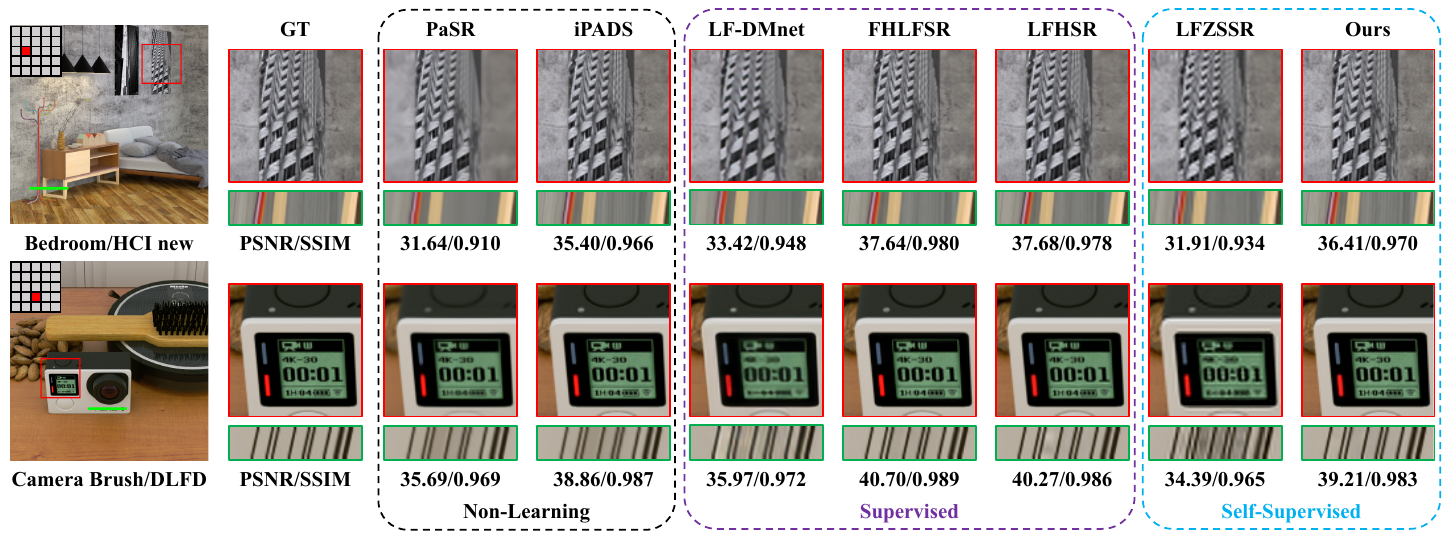}
    \vspace{-2.0em}
    \caption{Visual comparisons for $2\times$ SR on simulated dataset. The PSNR and SSIM scores achieved by different methods on the presented scenes are reported below the zoom-in regions, and the horizontal EPIs are shown. The colored grid on the top-left or top-right corner of each SAI indicates its angular position.}
    \label{fig:compareSimul}
\end{figure*}
\subsubsection{Comparisons of Visual Results} 
\begin{figure*}[htbp]
    \centering
    \includegraphics[width=\linewidth]{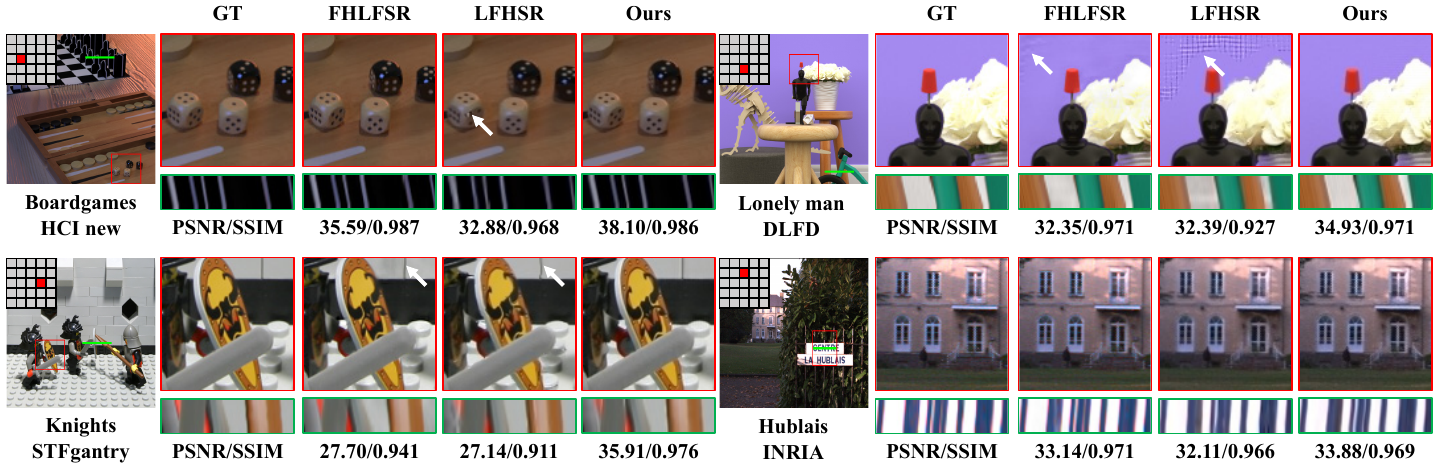}
    \vspace{-2.0em}
    \caption{Visual comparisons for $4\times$ SR results achieved by FHLFSR \cite{chang2022FHLFSR}, LFHSR \cite{jin2023LFHSR} and our method.}
    \label{fig:compareSimul_4}
\end{figure*}

Figures \ref{fig:compareSimul} and \ref{fig:compareSimul_4} illustrate the visual results for $2\times$ and $4\times$ SR tasks on simulated datasets, respectively. In every visual result for $2\times$ SR, a zoom-in region is given to illustrate the SR performance in detail, and an EPI from the representative region is given to illustrate the performance of maintaining the LF parallax structure. As shown in Figure \ref{fig:compareSimul}, when the domain gap is small, FHLFSR and LFHSR with hybrid inputs achieve relatively better visual results in all methods, and our self-supervised method achieves equally good SR visual results.
\par
In particular, in the $4\times$ SR task, since the domain gap between phases $b$ and $c$ is larger than that in the $2\times$ task, neither FHLFSR nor LFHSR can restore sufficient texture and structure information. Therefore, as shown in Figure \ref{fig:compareSimul_4}, the large domain gap results in serious artifacts in their SR results as denoted by white arrows. In contrast, our self-supervised method still successfully reconstructs high-frequency details and much clearer LF SAIs in $4\times$ SR task. 
\par
The visual results on simulated datasets demonstrate that our method reconstructs SR SAIs closer to the ground truth.
\subsubsection{Comparisons of the LF Parallax Structure} The LF parallax structure is the most valuable information in LF images compared with the traditional single images. As shown in Figures \ref{fig:compareSimul} and \ref{fig:compareSimul_4}, we selected the EPIs from representative areas of each scene for display. For the $2 \times$ SR task, the EPIs obtained by our method exhibited clear linear textures similar to those achieved by the SOTA supervised methods LFHSR \cite{jin2023LFHSR} and FHLFSR \cite{chang2022FHLFSR}. For the $4 \times$ SR task, as shown in Figure \ref{fig:compareSimul_4}, the large domain gap results in distorted lines in the EPIs obtained by LFHSR and FHLFSR. In contrast, our self-supervised method successfully maintains the LF parallax structure with much clearer linear textures in EPIs. To further evaluate the performance of different methods in preserving the LF parallax structure, we quantitatively compare the average SSIM of EPIs in $4\times$ SR task as previous works did \cite{jin2023LFHSR}. According to the quantitative comparison in Table \ref{tab:exp_comparison} and visual comparisons in Figure \ref{fig:compareSimul_4}, we only conduct comparisons with LFHSR and FHLFSR methods. As depicted in Table \ref{tab:exp_comparison3}, we achieve better SSIM scores compared with LFHSR and FHLFSR methods by $0.12$ and $0.26$, respectively. 
\par
Both the qualitative and quantitative results demonstrate that our SR method effectively preserves the LF parallax structure whether the domain gap is small or large. 
\subsection{Ablation study}
\label{subsec:ab}
We selectively remove the LF reorganization module (Framework 1), the EPI gradient loss (Framework 2), and the HR-aware loss (Framework 3) from the complete framework (Framework 4) to verify their effectiveness. 
\par
\begin{figure}[t]
    \centering
    \includegraphics[width=\linewidth]{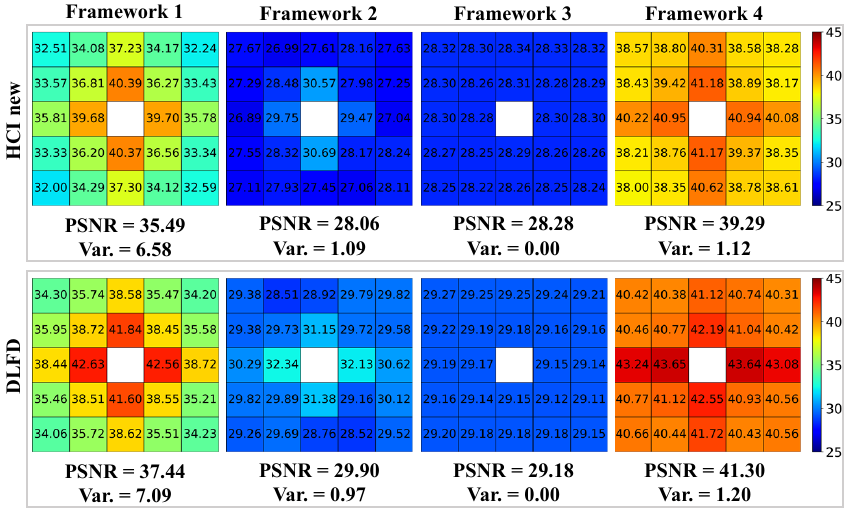}
    \vspace{-2.0em}
    \caption{Average PSNR distribution among different side-view SAIs obtained by different frameworks on HCI new and DLFD datasets. The average PSNR and the variances (Var.) in PSNR are listed below the heatmap.}
    \label{fig:compareAbVar}
\end{figure}

\textit{1) LF reorganization}: As shown in Table \ref{tab:ablation}, when we remove the LF reorganization module (Framework 1), the PSNR scores in both HCI new and DLFD datasets decrease by about $4$ dB compared with Framework 4. Furthermore, without the LF reorganization module, the CVS-Net tends to extract features from the side-view SAIs near the center due to information redundancy, so that the spatial SR performance of SAIs far from the center is lower than those close to the center, as shown in Figure \ref{fig:compareAbVar}. We use the variance of PSNR to verify the effectiveness of the LF reorganization module quantitatively. As shown in Figure \ref{fig:compareAbVar}, without the LF reorganization module (Framework 1), the variance increases significantly in both datasets. Therefore, we conclude that the LF reorganization module ensures that the network can extract features from all of the side-view SAIs evenly.
\par
\begin{figure}[!t]
    \centering
    \includegraphics[width=\linewidth]{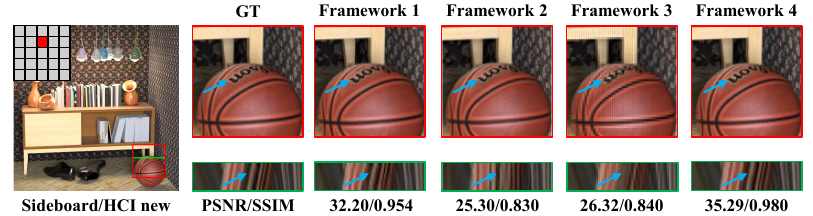}
    \vspace{-2.0em}
    \caption{Visual illustration for ablation study on loss functions. The horizontal EPIs, PSNR and SSIM scores are illustrated below the zoom-in regions.}
    \label{fig:compareAb}
\end{figure}

\textit{2) EPI gradient loss}: When the EPI gradient loss is not used to train HLFSSR-Net (Framework 2), the network loses the constraint for LF parallax structure, so that every SR side-view SAI will be similar to the only one HR central view reference image $I_{\mathbf{u}_0}^h$. As depicted in Table \ref{tab:ablation}, the PSNR on both datasets decreases over $10$ dB without the EPI gradient loss. More importantly, almost all of the lines in the EPI of Framework 2 are vertical, as shown in Figure \ref{fig:compareAb}. It means that the network without the EPI gradient loss fails to maintain the LF parallax structure. We refer readers to the supplementary file for more ablation study on EPI gradient loss.
\par
\textit{3) HR-aware loss}: As shown in Table \ref{tab:ablation}, the spatial SR performance decreases significantly when the HLFSSR-Net is supervised without the HR-aware loss (Framework 3), which is used to enhance the HLFSSR-Net's ability in detail recovery. Furthermore, as shown in Figure \ref{fig:compareAb} (Framework 3), the network produces seriously blurry results and fails to recover the texture details. It should be noted that we remove the LF reorganization module when we remove the HR-aware loss, as shown in Table \ref{tab:ablation}, because the LF reorganization module is useless without the HR-aware loss. 
\par
In summary, the results of the ablation study of loss functions illustrate the indispensable importance of the two loss functions and the LF reorganization module.
\par
\par
\begin{table}[!t]
\centering
\renewcommand\arraystretch{1.25}
    \caption{Ablation study on loss functions. The PSNR/SSIM scores of $2\times$ SR results on $5 \times 5$ LFs are provided. $\ell_{epi}$: EPI Gradient Loss, $ \ell_{hr} $: HR-aware Loss, R: LF Reorganization.}
    \label{tab:ablation}
    \vspace{-0.3cm}
\resizebox{\linewidth}{!}
{
    \centering
    \begin{tabular}{c|c|c|c|c|c|c|c}
    \toprule
         \multirow{2}*{Framework} & \multicolumn{3}{c|}{Elements} & \multicolumn{2}{c|}{\#Param.} &\multicolumn{2}{c}{PSNR/SSIM} \\
         \cline{2-8}
         ~ & $\ell_{epi}$ & $ \ell_{hr} $ & R & CVS-Net & BD-Net & HCI new & DLFD   \\
         \hline
          1 & $\checkmark$ & $\checkmark$  & -  & 7.12M & 0.37M & 35.49/0.965 &	37.44/0.972 \\ 
          2 & - & $\checkmark$  & $\checkmark$  & 2.71M & - & 28.06/0.880 &	29.90/0.899 \\
          3 & $\checkmark$ & -  & -  & - & 0.37M & 28.28/0.852 &	29.18/0.842\\
          4 & $\checkmark$ & $\checkmark$ & $\checkmark$ & 2.71M & 0.37M & \textbf{39.29}/\textbf{0.984} &	\textbf{41.30}/\textbf{0.987} \\
         \bottomrule
    \end{tabular}
    }
\end{table}
\par

\section{Experiments on Our Real-world Hybrid Dataset}
\label{sec:exp_real}
\subsection{Implementation Details}
To demonstrate the advantage of our self-supervised method, we conduct further evaluations on our real-world hybrid dataset described in Section \ref{sec:system}. We apply the same pre-processing operations to the images in our hybrid dataset, as depicted in Sub-section \ref{subsec:implement}. We adopt Adam optimizer \cite{kingma2014adam} with $\beta_{1} = 0.9$, $\beta_{2} = 0.999$ for both the pre-training and training stages. The learning rate of all networks is set to $2\times10^{-4}$ initially and decreases by a factor of 0.5 every 15 epochs for training CVS-Net and HLFSSR-Net, and decreases by a factor of 0.5 every 125 epochs for training BD-Net. Due to the complexity of the image degradation on our real-world dataset, the training of the BD-Net is stopped after 500 epochs. The training of both the CVS-Net and the HLFSSR-Net is stopped after 50 epochs. To retrain the compared supervised networks \cite{wang2024real-worldLF,jin2023LFHSR,chang2022FHLFSR} on our real-world hybrid LF dataset, as the ground truth (HR LF SAIs) is not available for supervision, we follow Sub-section \ref{subsec:exp_setup} to simulate training datasets and follow the implementation details described in these paper \cite{wang2024real-worldLF,jin2023LFHSR,chang2022FHLFSR} to ensure fair comparisons.

\subsection{Comparisons with State-of-the-art Methods}
As there is no HR LF ground truth in our real-world hybrid dataset, we perform a no-reference image quality assessment using the BRISQUE (Blind/Referenceless Image Spatial Quality Evaluator) metric \cite{Mittal2012BRISQUE}, and compare the visual SR results obtained by SOTA methods and our method. 
We refer readers to the supplementary file for estimated disparity maps.
\par
\begin{figure*}[htbp]
    \centering
    \includegraphics[width=0.9\linewidth]{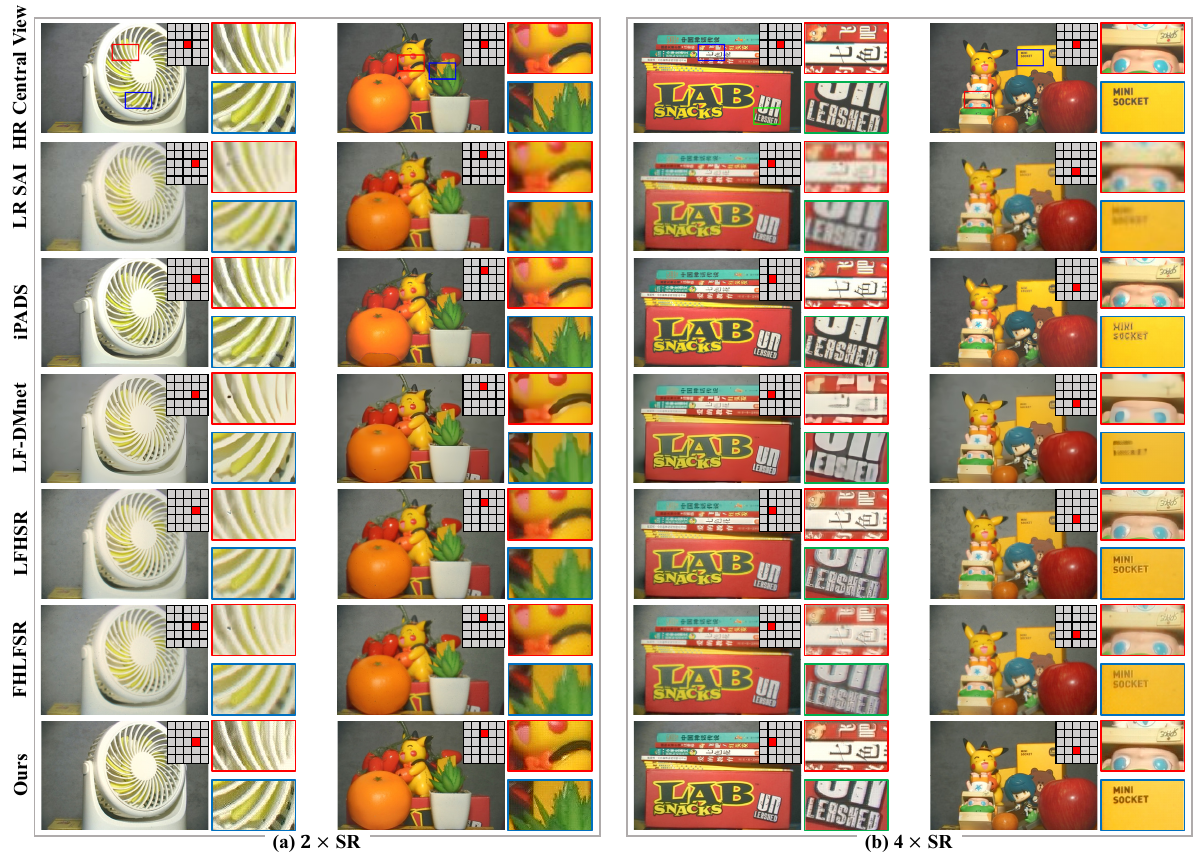}
    \vspace{-1.0em}
    \caption{Comparisons of visual results for (a) $2\times$ SR and (b) $4\times$ SR on our real-world hybrid LF data. 
    % The colored grid on the top-right corner of each SAI indicates its angular position.
    }
    \label{fig:compareOur}
\end{figure*}

\begin{table}[!ht]
    \centering
    \caption{BRISQUE($\downarrow$) \cite{Mittal2012BRISQUE} results achieved by HR center view, LR SAIs, and different methods (iPADs\cite{wang2016iPADs}, LF-DMnet\cite{wang2024real-worldLF}, LFHSR\cite{jin2023LFHSR}, FHLFSR\cite{chang2022FHLFSR}, and Ours) on real-world hybrid LF dataset for $2 \times$SR and $4 \times$SR.}
    \vspace{-0.3cm}
    \resizebox{\linewidth}{!}
    {
    \begin{tabular}{c|ccccccc}
    \toprule
         Method & HR & LR & \cite{wang2016iPADs} & \cite{wang2024real-worldLF} & \cite{jin2023LFHSR} & \cite{chang2022FHLFSR} & Ours\\
         \hline
         $2 \times$SR & 10.245 & 32.856 & 29.797 & 48.031 & 32.795 & 35.924 & 13.338 \\
         \hline
         $4 \times$SR & 19.681 & 32.856 & 42.525 & 72.597 & 36.229 & 50.357 & 20.741 \\
         \bottomrule
         
    \end{tabular}
    }
    \label{tab:brisque}
\end{table}
As shown in Table \ref{tab:brisque}, our self-supervised method achieves BRISQUE scores of approximately 13 for $2 \times$ SR and 20 for $4 \times$ SR---both within the high-quality range and close to the HR center-view scores ($<20$). In contrast, the SOTA methods \cite{wang2016iPADs,wang2024real-worldLF, jin2023LFHSR,chang2022FHLFSR} yield noticeably higher BRISQUE values ($>20$), indicating stronger distortion (as observable in Figure \ref{fig:compareOur}). These results quantitatively confirm that our proposed method produces superior real-world reconstruction quality.
\par
Moreover, as shown in Figure \ref{fig:compareOur}(a), our method reconstructs much sharper edges and clearer scenes than all compared methods including the FHLFSR \cite{chang2022FHLFSR} and LFHSR \cite{jin2023LFHSR}, which achieved higher PSNR/SSIM scores than our method in $2\times$ SR tasks on simulated datasets. Compared with the simulated datasets, there exists a larger domain gap between phases $b$ and $c$ in our real-world hybrid dataset, which cannot be handled well by the FHLFSR and LFHSR. For $4\times$ SR results, as shown in zoom-in regions of Figure \ref{fig:compareOur}(b), all compared methods except ours suffer from serious blur and distortion, such as the Chinese and English letters on the books and boxes. Although LFHSR achieves acceptable performance compared with other SOTA methods, our approach keeps much better high-frequency details that are similar to the 2D HR image.
In addition, as shown in Figure \ref{fig:compareOur}, the simple yet effective method LF-DMnet \cite{wang2024real-worldLF} for real-world LF image SR, produces images with artifacts or over-smoothness when the input blur kernel widths and noise levels during training mismatch with the real ones during inference, this phenomenon is as described in the paper \cite{wang2024real-worldLF}. It's difficult to precisely measure or estimate the PSF and the noise level of a real-world camera and generalize them to real-world applications in a short period.
\par
In summary, as our self-supervised method addresses the issue of LF spatial SR without the pre-defined degradation models, it produces satisfactory results in both $2\times$ and $4\times$ SR tasks on our real-world hybrid dataset. Therefore, these results further demonstrate the significant advantages of our method over the SOTA ones.

\section{Conclusion}
\label{sec:conclusion}
When existing learning-based methods are used to produce spatial super-resolved LF images whose resolution is higher than that of the originally captured LF images, they inevitably face the domain gap due to the lack of HR ground truth. To address this issue, this paper proposes a novel solution that consists of three correlative parts. The foundation of our work is the hybrid LF imaging prototype, which simultaneously captures the 4D LR LF image and a corresponding central view 2D HR reference image. We capture 189 scenes using this prototype to create a hybrid dataset tailored for training our proposed self-supervised LF spatial SR framework. According to the characteristics of our solution, the self-supervised framework with HLFSSR-Net, CVS-Net, and BD-Net is well-designed to propagate high-frequency information from only one HR image to all side-view SAIs and preserve LF parallax structures. Our method essentially overcomes the domain gap that is commonly faced in supervised approaches, and extensive experimental results on both simulated datasets and our real-world hybrid dataset demonstrate the effectiveness and generalization of our method in comparison with SOTA methods. The results indicate that our approach not only meets but exceeds the capabilities of existing supervised solutions, paving the way for broader adoption in real-world applications.

% or


% \appendices

% \section{}

% \section*{Acknowledgment}


\ifCLASSOPTIONcaptionsoff
  \newpage
\fi

\normalem

\bibliographystyle{IEEEtran}
\bibliography{bibtex/bib/IEEEabrv,bibtex/bib/bib}

@misc{Lytro,
   author = {Lytro},
   title = {Lytro redefines photography with light field cameras},
   howpublished = {\url{http://lightfield.stanford.edu/acq.html}},      
   year={2011}
}

@inproceedings{Levoy1996LFRendering,
author = {Levoy, Marc and Hanrahan, Pat},
title = {Light field rendering},
year = {1996},
isbn = {0897917464},
publisher = {Association for Computing Machinery},
address = {New York, NY, USA},
booktitle = {Proceedings of the 23rd Annual Conference on Computer Graphics and Interactive Techniques},
pages = {31–42},
numpages = {12},
series = {SIGGRAPH '96}
}

@ARTICLE{zhao2024refocusing,
  author={Zhao, Chun and Jeon, Byeungwoo},
  journal={IEEE Transactions on Computational Imaging}, 
  title={Compact Representation of Light Field Data for Refocusing and Focal Stack Reconstruction Using Depth Adaptive Multi-CNN}, 
  year={2024},
  volume={10},
  number={},
  pages={170-180},
  doi={10.1109/TCI.2023.3347910}}

@article{yu2017vr,
  title={A light-field journey to virtual reality},
  author={Yu, Jingyi},
  journal={IEEE MultiMedia},
  volume={24},
  number={2},
  pages={104--112},
  year={2017}
}

@article{cai2018ray,
  title={Ray calibration and phase mapping for structured-light-field 3D reconstruction},
  author={Cai, Zewei and Liu, Xiaoli and Peng, Xiang and Gao, Bruce Z},
  journal={Optics Express},
  volume={26},
  number={6},
  pages={7598--7613},
  year={2018},
  publisher={Optica Publishing Group}
}

@article{Zhou20223d,
  author = {Ping Zhou and Yanzheng Wang and Yuda Xu and Zewei Cai and Chao Zuo},
  journal = {Optics Express},
  number = {17},
  pages = {29957--29968},
  publisher = {Optica Publishing Group},
  title = {Phase-unwrapping-free 3D reconstruction in structured light field system based on varied auxiliary point},
  volume = {30},
  year = {2022},
}

@article{zhou2019calibration,
title = {A two-step calibration method of lenslet-based light field cameras},
journal = {Optics and Lasers in Engineering},
volume = {115},
pages = {190-196},
year = {2019},
issn = {0143-8166},
author = {Ping Zhou and Weijia Cai and Yunlei Yu and Yuting Zhang and Guangquan Zhou},

}

@INPROCEEDINGS{zhangzhengyou,
  author={Zhengyou Zhang},
  booktitle={Proceedings of the Seventh IEEE International Conference on Computer Vision}, 
  title={Flexible camera calibration by viewing a plane from unknown orientations}, 
  year={1999},
  volume={1},
  number={},
  pages={666-673},
  
}

@INPROCEEDINGS{xiong2017hybrid,
  author={Xiong, Zhiwei and Wang, Lizhi and Li, Huiqun and Liu, Dong and Wu, Feng},
  booktitle={2017 IEEE Conference on Computer Vision and Pattern Recognition (CVPR)}, 
  title={Snapshot Hyperspectral Light Field Imaging}, 
  year={2017},
  volume={},
  number={},
  pages={6873-6881},

}

@inproceedings{boominathan2014PaSR,
  title={Improving resolution and depth-of-field of light field cameras using a hybrid imaging system},
  author={Boominathan, Vivek and Mitra, Kaushik and Veeraraghavan, Ashok},
  booktitle={2014 IEEE International Conference on Computational Photography (ICCP)},
  pages={1--10},
  year={2014},
}

@article{alam2018hybrid,
  title={Hybrid light field imaging for improved spatial resolution and depth range},
  author={Alam, M Zeshan and Gunturk, Bahadir K},
  journal={Machine Vision and Applications},
  volume={29},
  number={1},
  pages={11--22},
  year={2018},
  publisher={Springer}
}

@article{wang2016iPADs,
  title={The light field attachment: Turning a DSLR into a light field camera using a low budget camera ring},
  author={Wang, Yuwang and Liu, Yebin and Heidrich, Wolfgang and Dai, Qionghai},
  journal={IEEE transactions on visualization and computer graphics},
  volume={23},
  number={10},
  pages={2357--2364},
  year={2016},
  publisher={IEEE}
}

@article{wang2016hrHybrid,
  author = {Xiang Wang and Lin Li and GuangQi Hou},
  journal = {Appl. Opt.},
  number = {10},
  pages = {2580--2593},
  publisher = {Optica Publishing Group},
  title = {High-resolution light field reconstruction using a hybrid imaging system},
  volume = {55},
  year = {2016},


}

@article{wang2017lfvideoHybrid,
  author = {Wang, Ting-Chun and Zhu, Jun-Yan and Kalantari, Nima Khademi and Efros, Alexei A. and Ramamoorthi, Ravi},
  title = {Light Field Video Capture Using a Learning-Based Hybrid Imaging System},
  year = {2017},
  issue_date = {August 2017},
  publisher = {Association for Computing Machinery},
  address = {New York, NY, USA},
  volume = {36},
  number = {4},
  issn = {0730-0301},
  journal = {ACM Trans. Graph.},
  articleno = {133},
  numpages = {13},
}

@article{yang2023Hybridlenses,
  author={Yang, Yang and Wu, Lianxiong and Zeng, Lanling and Yan, Tao and Zhan, Yongzhao},
  journal={IEEE Transactions on Instrumentation and Measurement}, 
  title={Joint Upsampling for Refocusing Light Fields Derived With Hybrid Lenses}, 
  year={2023},
  volume={72},
  number={5009512},
 pages={1-12},

}

@article{ye2023LFIENet,
  author={Ye, Wuyang and Yan, Tao and Gao, Jiahui and Yang, Yang},
  journal={IEEE Transactions on Computational Imaging}, 
  title={LFIENet: Light Field Image Enhancement Network by Fusing Exposures of LF-DSLR Image Pairs}, 
  year={2023},
  volume={9},
  number={},
  pages={620-635},

}

@ARTICLE{Guo2022DRLF,
  author={Guo, Mantang and Hou, Junhui and Jin, Jing and Chen, Jie and Chau, Lap-Pui},
  journal={IEEE Transactions on Pattern Analysis and Machine Intelligence}, 
  title={Deep Spatial-Angular Regularization for Light Field Imaging, Denoising, and Super-Resolution}, 
  year={2022},
  volume={44},
  number={10},
  pages={6094-6110},
  }

@article{Bishop2012,
  author={Bishop, Tom E. and Favaro, Paolo},
  journal={IEEE Transactions on Pattern Analysis and Machine Intelligence}, 
  title={The Light Field Camera: Extended Depth of Field, Aliasing, and Superresolution}, 
  year={2012},
  volume={34},
  number={5},
  pages={972-986},
  }

@article{Wanner2014,
  author={Wanner, Sven and Goldluecke, Bastian},
  journal={IEEE Transactions on Pattern Analysis and Machine Intelligence}, 
  title={Variational Light Field Analysis for Disparity Estimation and Super-Resolution}, 
  year={2014},
  volume={36},
  number={3},
  pages={606-619},
 }

@article{Farrugia2017,
  author={Farrugia, Reuben A. and Galea, Christian and Guillemot, Christine},
  journal={IEEE Journal of Selected Topics in Signal Processing}, 
  title={Super Resolution of Light Field Images Using Linear Subspace Projection of Patch-Volumes}, 
  year={2017},
  volume={11},
  number={7},
  pages={1058-1071},
 }

@inproceedings{Alain2018,
  author={Alain, Martin and Smolic, Aljosa},
  booktitle={2018 25th IEEE International Conference on Image Processing (ICIP)}, 
  title={Light Field Super-Resolution via LFBM5D Sparse Coding}, 
  year={2018},
  volume={},
  number={},
  pages={2501-2505},
 }

@ARTICLE{Rossi2018,
  author={Rossi, Mattia and Frossard, Pascal},
  journal={IEEE Transactions on Image Processing}, 
  title={Geometry-Consistent Light Field Super-Resolution via Graph-Based Regularization}, 
  year={2018},
  volume={27},
  number={9},
  pages={4207-4218},
 }

@inproceedings{cheng2021LFZSSR,
  author    = {Cheng, Zhen and Xiong, Zhiwei and Chen, Chang and Liu, Dong and Zha, Zheng-Jun},
  title     = {Light Field Super-Resolution With Zero-Shot Learning},
  booktitle = {Proceedings of the IEEE/CVF Conference on Computer Vision and Pattern Recognition (CVPR)},
  year      = {2021},
  pages     = {10010-10019}
}

@ARTICLE{sheng2023lfassr,
  author={Sheng, Hao and Wang, Sizhe and Yang, Da and Cong, Ruixuan and Cui, Zhenglong and Chen, Rongshan},
  journal={IEEE Transactions on Circuits and Systems for Video Technology}, 
  title={Cross-View Recurrence-based Self-Supervised Super-Resolution of Light Field}, 
  year={2023},
  volume={33},
  number={12},
  pages={7252-7266},

}

@inproceedings{jin2020AlltoOne,
  author = {Jin, Jing and Hou, Junhui and Chen, Jie and Kwong, Sam},
  title = {Light Field Spatial Super-Resolution via Deep Combinatorial Geometry Embedding and Structural Consistency Regularization},
  booktitle = {Proceedings of the IEEE/CVF Conference on Computer Vision and Pattern Recognition (CVPR)},
  year = {2020},
  pages={2257-2266},
}

@article{liang2022LFT,
  author={Liang, Zhengyu and Wang, Yingqian and Wang, Longguang and Yang, Jungang and Zhou, Shilin},
  journal={IEEE Signal Processing Letters}, 
  title={Light Field Image Super-Resolution With Transformers}, 
  year={2022},
  volume={29},
  number={},
  pages={563-567},

}

@inproceedings{liang2023EPIT,
  author    = {Liang, Zhengyu and Wang, Yingqian and Wang, Longguang and Yang, Jungang and Zhou, Shilin and Guo, Yulan},
  title     = {Learning Non-Local Spatial-Angular Correlation for Light Field Image Super-Resolution},
  booktitle = {Proceedings of the IEEE/CVF International Conference on Computer Vision (ICCV)},
  year      = {2023},
  pages     = {12376--12386},
}

@article{wang2018LFNet,
  author={Wang, Yunlong and Liu, Fei and Zhang, Kunbo and Hou, Guangqi and Sun, Zhenan and Tan, Tieniu},
  journal={IEEE Transactions on Image Processing}, 
  title={LFNet: A Novel Bidirectional Recurrent Convolutional Neural Network for Light-Field Image Super-Resolution}, 
  year={2018},
  volume={27},
  number={9},
  pages={4274-4286},

}

@article{wang2022disentangling,
  title={Disentangling light fields for super-resolution and disparity estimation},
  author={Wang, Yingqian and Wang, Longguang and Wu, Gaochang and Yang, Jungang and An, Wei and Yu, Jingyi and Guo, Yulan},
  journal={IEEE Transactions on Pattern Analysis and Machine Intelligence},
  volume={45},
  number={1},
  pages={425--443},
  year={2022},
  publisher={IEEE}
}

@article{yeung2018LFSSR,
  title={Light field spatial super-resolution using deep efficient spatial-angular separable convolution},
  author={Yeung, Henry Wing Fung and Hou, Junhui and Chen, Xiaoming and Chen, Jie and Chen, Zhibo and Chung, Yuk Ying},
  journal={IEEE Transactions on Image Processing},
  volume={28},
  number={5},
  pages={2319--2330},
  year={2018},
  publisher={IEEE}
}

@inproceedings{yoon2015LFSR,
  author = {Yoon, Youngjin and Jeon, Hae-Gon and Yoo, Donggeun and Lee, Joon-Young and So Kweon, In},
  title = {Learning a Deep Convolutional Network for Light-Field Image Super-Resolution},
  booktitle = {Proceedings of the IEEE International Conference on Computer Vision (ICCV) Workshops},
  year = {2015},
  pages={57-65},
}

@article{yuan2018LFSREPI,
  author={Yuan, Yan and Cao, Ziqi and Su, Lijuan},
  journal={IEEE Signal Processing Letters}, 
  title={Light-Field Image Superresolution Using a Combined Deep CNN Based on EPI}, 
  year={2018},
  volume={25},
  number={9},
  pages={1359-1363},

}

@inproceedings{zhang2019resLF,
  author = {Zhang, Shuo and Lin, Youfang and Sheng, Hao},
  title = {Residual Networks for Light Field Image Super-Resolution},
  booktitle = {Proceedings of the IEEE/CVF Conference on Computer Vision and Pattern Recognition (CVPR)},
  year = {2019},
  pages={11038-11047},
}

@ARTICLE{wang2024real-worldLF,
  author={Wang, Yingqian and Liang, Zhengyu and Wang, Longguang and Yang, Jungang and An, Wei and Guo, Yulan},
  journal={IEEE Transactions on Neural Networks and Learning Systems}, 
  title={Real-World Light Field Image Super-Resolution Via Degradation Modulation}, 
  year={2025},
  volume={36},
  number={3},
  pages={5559-5573},

}

@InProceedings{xiao2023real-worldLF,
    author    = {Xiao, Zeyu and Gao, Ruisheng and Liu, Yutong and Zhang, Yueyi and Xiong, Zhiwei},
    title     = {Toward Real-World Light Field Super-Resolution},
    booktitle = {Proceedings of the IEEE/CVF Conference on Computer Vision and Pattern Recognition (CVPR) Workshops},
    month     = {June},
    year      = {2023},
    pages     = {3408-3418}
}

@article{lyu2024probabilistic,
  title={Probabilistic-based feature embedding of 4-d light fields for compressive imaging and denoising},
  author={Lyu, Xianqiang and Hou, Junhui},
  journal={International Journal of Computer Vision},
  volume={132},
  number={6},
  pages={2255--2275},
  year={2024},
  publisher={Springer}
}

@InProceedings{Chen_2024_CVPR,
    author    = {Chen, Haoyu and Li, Wenbo and Gu, Jinjin and Ren, Jingjing and Sun, Haoze and Zou, Xueyi and Zhang, Zhensong and Yan, Youliang and Zhu, Lei},
    title     = {Low-Res Leads the Way: Improving Generalization for Super-Resolution by Self-Supervised Learning},
    booktitle = {Proceedings of the IEEE/CVF Conference on Computer Vision and Pattern Recognition (CVPR)},
    month     = {June},
    year      = {2024},
    pages     = {25857-25867}
}

@article{chen2023LFSSR-HI,
  author={Chen, Yeyao and Jiang, Gangyi and Yu, Mei and Xu, Haiyong and Ho, Yo-Sung},
  journal={IEEE Transactions on Visualization and Computer Graphics}, 
  title={Deep Light Field Spatial Super-Resolution Using Heterogeneous Imaging}, 
  year={2023},
  volume={29},
  pages={4183-4197},

}

@inproceedings{jin2020LFHSR,
  author = {Jin, Jing and Hou, Junhui and Chen, Jie and Kwong, Sam and Yu, Jingyi},
  title = {Light Field Super-Resolution via Attention-Guided Fusion of Hybrid Lenses},
  year = {2020},
  isbn = {9781450379885},
  publisher = {Association for Computing Machinery},
  address = {New York, NY, USA},  
  booktitle = {Proceedings of the 28th ACM International Conference on Multimedia},
  pages = {193–201},
  numpages = {9},
  location = {Seattle, WA, USA},
  series = {MM '20},

}

@article{jin2023LFHSR,
  title={Light Field Reconstruction Via Deep Adaptive Fusion of Hybrid Lenses},
  author={Jin, Jing and Guo, Mantang and Hou, Junhui and Liu, Hui and Xiong, Hongkai},
  journal={IEEE Transactions on Pattern Analysis and Machine Intelligence},
  year={2023},
  volume={45},

  pages={12050-12067},
  publisher={IEEE}
}

@article{zhao2018HCSR,
  author={Zhao, Mandan and Wu, Gaochang and Li, Yipeng and Hao, Xiangyang and Fang, Lu and Liu, Yebin},
  journal={IEEE Transactions on Computational Imaging}, 
  title={Cross-Scale Reference-Based Light Field Super-Resolution}, 
  year={2018},
  volume={4},
  number={3},
  pages={406-418},

}

@inproceedings{zheng2017hybrid,
author = {Zheng, Haitian and Guo, Minghao and Wang, Haoqian and Liu, Yebin and Fang, Lu},
title = {Combining Exemplar-Based Approach and Learning-Based Approach for Light Field Super-Resolution Using a Hybrid Imaging System},
booktitle = {Proceedings of the IEEE International Conference on Computer Vision (ICCV) Workshops},
year = {2017},
pages={2481-2486},
}

@inproceedings{chang2022FHLFSR,
author = {Chang, Song and Lin, Youfang and Zhang, Shuo},
title = {Flexible Hybrid Lenses Light Field Super-Resolution Using Layered Refinement},
year = {2022},
isbn = {9781450392037},
publisher = {Association for Computing Machinery},
address = {New York, NY, USA},
booktitle = {Proceedings of the 30th ACM International Conference on Multimedia},
pages = {5584–5592},
numpages = {9},
location = {Lisboa, Portugal},
series = {MM '22}
}

@inproceedings{honauer2017HCI,
 author="Honauer, Katrin and Johannsen, Ole and Kondermann, Daniel and Goldluecke, Bastian", 
 title="A Dataset and Evaluation Methodology for Depth Estimation on 4D Light Fields",
 booktitle="Asian Conference on Computer Vision (ACCV)",
 year="2017", 
 pages="19--34",
}

@article{shi2019DLFD,
  title={A framework for learning depth from a flexible subset of dense and sparse light field views},
  author={Shi, Jinglei and Jiang, Xiaoran and Guillemot, Christine},
  journal={IEEE Transactions on Image Processing},
  volume={28},
  number={12},
  pages={5867--5880},
  year={2019},
  publisher={IEEE}
}

@inproceedings{rerabek2016EPFL,
  title={New light field image dataset},
  author={Rerabek, Martin and Ebrahimi, Touradj},
  booktitle={8th International Conference on Quality of Multimedia Experience (QoMEX)},
  number={CONF},
  year={2016}
}

@article{le2018INRIA,
  title={Light field inpainting propagation via low rank matrix completion},
  author={Le Pendu, Mikael and Jiang, Xiaoran and Guillemot, Christine},
  journal={IEEE Transactions on Image Processing},
  volume={27},
  number={4},
  pages={1981--1993},
  year={2018},
  publisher={IEEE}
}

@misc{vaish2008STFgantry,
  title = {The (New) Stanford Light Field Archive},
  howpublished = "\url{http://lightfield.stanford.edu/lfs.html}",
  author = {Vaibhav Vaish and Andrew Adams},
  year = {2008},
  note = "[Online]"
}

@article{srivastava2014dropout,
  title={Dropout: a simple way to prevent neural networks from overfitting},
  author={Srivastava, Nitish and Hinton, Geoffrey and Krizhevsky, Alex and Sutskever, Ilya and Salakhutdinov, Ruslan},
  journal={The journal of machine learning research},
  volume={15},
  number={1},
  pages={1929--1958},
  year={2014},
  publisher={JMLR. org}
}

@inproceedings{hu2018CAM,
author = {Hu, Jie and Shen, Li and Sun, Gang},
title = {Squeeze-and-Excitation Networks},
booktitle = {Proceedings of the IEEE Conference on Computer Vision and Pattern Recognition (CVPR)},
month = {June},
year = {2018},
pages={7132-7141},
}

@InProceedings{shi2016pixelShuffle,
author = {Shi, Wenzhe and Caballero, Jose and Huszar, Ferenc and Totz, Johannes and Aitken, Andrew P. and Bishop, Rob and Rueckert, Daniel and Wang, Zehan},
title = {Real-Time Single Image and Video Super-Resolution Using an Efficient Sub-Pixel Convolutional Neural Network},
booktitle = {Proceedings of the IEEE Conference on Computer Vision and Pattern Recognition (CVPR)},
month = {June},
year = {2016},
pages={1874-1883},
}

@inproceedings{guo2019CBDNet,
  author = {Guo, Shi and Yan, Zifei and Zhang, Kai and Zuo, Wangmeng and Zhang, Lei},
  title = {Toward Convolutional Blind Denoising of Real Photographs},
  booktitle = {Proceedings of the IEEE/CVF Conference on Computer Vision and Pattern Recognition (CVPR)},
  year = {2019},
  pages={1712-1722},
}

@inproceedings{jin2020LFASR,
  title={Learning light field angular super-resolution via a geometry-aware network},
  author={Jin, Jing and Hou, Junhui and Yuan, Hui and Kwong, Sam},
  booktitle={Proceedings of the AAAI conference on artificial intelligence},
  volume={34},
  number={07},
  pages={11141-11148},
  year={2020}
}

@ARTICLE{jin2022LFASR,
  author={Jin, Jing and Hou, Junhui and Chen, Jie and Zeng, Huanqiang and Kwong, Sam and Yu, Jingyi},
  journal={IEEE Transactions on Pattern Analysis and Machine Intelligence}, 
  title={Deep Coarse-to-Fine Dense Light Field Reconstruction With Flexible Sampling and Geometry-Aware Fusion}, 
  year={2022},
  volume={44},
  number={4},
  pages={1819-1836},
  doi={10.1109/TPAMI.2020.3026039}}

@article{kingma2014adam,
  title={Adam: A method for stochastic optimization},
  author={Kingma, Diederik P and Ba, Jimmy},
  journal={arXiv preprint arXiv:1412.6980},
  year={2014}
}

@ARTICLE{liu2023BlindSR,
  author={Liu, Anran and Liu, Yihao and Gu, Jinjin and Qiao, Yu and Dong, Chao},
  journal={IEEE Transactions on Pattern Analysis and Machine Intelligence}, 
  title={Blind Image Super-Resolution: A Survey and Beyond}, 
  year={2023},
  volume={45},
  number={5},
  pages={5461-5480},
  doi={10.1109/TPAMI.2022.3203009}}

@InProceedings{Wei2021DASR,
    author    = {Wei, Yunxuan and Gu, Shuhang and Li, Yawei and Timofte, Radu and Jin, Longcun and Song, Hengjie},
    title     = {Unsupervised Real-World Image Super Resolution via Domain-Distance Aware Training},
    booktitle = {Proceedings of the IEEE/CVF Conference on Computer Vision and Pattern Recognition (CVPR)},
    month     = {June},
    year      = {2021},
    pages     = {13385-13394}
}

@inproceedings{yuan2018CinCGAN,
  author = {Yuan, Yuan and Liu, Siyuan and Zhang, Jiawei and Zhang, Yongbing and Dong, Chao and Lin, Liang},
  title = {Unsupervised Image Super-Resolution Using Cycle-in-Cycle Generative Adversarial Networks},
  booktitle = {Proceedings of the IEEE Conference on Computer Vision and Pattern Recognition (CVPR) Workshops},
  year = {2018},
  pages = {814-823},
}

@inproceedings{ashraf2018cc,
  author={Ashraf, Maliha and Sapaico, Luis Ricardo},
  booktitle={2018 Colour and Visual Computing Symposium (CVCS)}, 
  title={Evaluation of Color Correction Methods for Printed Surfaces}, 
  year={2018},
  volume={},
  number={},
  pages={1-6},
}

@inproceedings{zhang2018LPIPS,
  author = {Zhang, Richard and Isola, Phillip and Efros, Alexei A. and Shechtman, Eli and Wang, Oliver},
  title = {The Unreasonable Effectiveness of Deep Features as a Perceptual Metric},
  booktitle = {Proceedings of the IEEE Conference on Computer Vision and Pattern Recognition (CVPR)},
  year = {2018},
  pages={586-595},
}

@ARTICLE{Mittal2012BRISQUE,
  author={Mittal, Anish and Moorthy, Anush Krishna and Bovik, Alan Conrad},
  journal={IEEE Transactions on Image Processing}, 
  title={No-Reference Image Quality Assessment in the Spatial Domain}, 
  year={2012},
  volume={21},
  number={12},
  pages={4695-4708},
  doi={10.1109/TIP.2012.2214050}}

@article{chen2018depth,
  author={Chen, Jie and Hou, Junhui and Ni, Yun and Chau, Lap-Pui},
  journal={IEEE Transactions on Image Processing}, 
  title={Accurate Light Field Depth Estimation With Superpixel Regularization Over Partially Occluded Regions}, 
  year={2018},
  volume={27},
  number={10},
  pages={4889-4900},
}

@article{jin2022occlusion,
  title={Occlusion-aware unsupervised learning of depth from 4-D light fields},
  author={Jin, Jing and Hou, Junhui},
  journal={IEEE Transactions on Image Processing},
  volume={31},
  pages={2216--2228},
  year={2022},
  publisher={IEEE}
}

@article{zhou2023light,
  title={Light field depth estimation via stitched epipolar plane images},
  author={Zhou, Ping and Shi, Langqing and Liu, Xiaoyang and Jin, Jing and Zhang, Yuting and Hou, Junhui},
  journal={IEEE Transactions on Visualization and Computer Graphics},
  volume={30},
  number={10},
  pages={6866--6879},
  year={2023},
  publisher={IEEE}
}

@article{Gui2024SSL,
  author={Gui, Jie and Chen, Tuo and Zhang, Jing and Cao, Qiong and Sun, Zhenan and Luo, Hao and Tao, Dacheng},
  journal={IEEE Transactions on Pattern Analysis and Machine Intelligence}, 
  title={A Survey on Self-Supervised Learning: Algorithms, Applications, and Future Trends}, 
  year={2024},
  volume={46},
  number={12},
  pages={9052-9071},
  doi={10.1109/TPAMI.2024.3415112}}

\end{document}

\title{Self-supervised Learning-based Reconstruction of High-resolution 4D Light Fields \\ (\textit{Supplementary Materials})}

\author{Jianxin~Lei,
        Dongze~Wu,
        Chengcai~Xu,
        Hongcheng~Gu,
        Guangquan~Zhou,~\IEEEmembership{Member,~IEEE},
        Junhui~Hou,~\IEEEmembership{Senior~Member,~IEEE},
        and~Ping~Zhou,~\IEEEmembership{Member,~IEEE}}% <-this % stops a space

% \markboth{Journal of \LaTeX\ Class Files,~Vol.~14, No.~8, August~2015}%
% \markboth{MANUSCRIPT SUBMITTED TO IEEE TRANSACTIONS ON VISUALIZATION AND COMPUTER GRAPHICS}
% {Shell \MakeLowercase{\textit{et al.}}: Bare Demo of IEEEtran.cls for IEEE Journals}

\maketitle
In this Supplementary Material, we present additional details supporting the main manuscript. Section~\ref{sec:s1} describes the detailed network architectures of the proposed HLFSSR-Net and CVS-Net. Section~\ref{sec:s2} provides further illustration of the LF reorganization strategies and reports the average PSNR distributions across different angular resolutions. Section~\ref{sec:s3} presents the ablation study on the EPI gradient loss. Section~\ref{sec:s4} includes the comparisons of disparity maps estimated from the LR LF images and the super-resolved LF images produced by different methods on our real-world hybrid dataset. Finally, Section~\ref{sec:s5} reports concise yet compelling quantitative results to demonstrate the mitigation of the domain gap achieved by our approach.
\par
\section{network architecture}
\label{sec:s1}
\begin{figure*}[htbp]
    \centering
    \includegraphics[width=\linewidth]{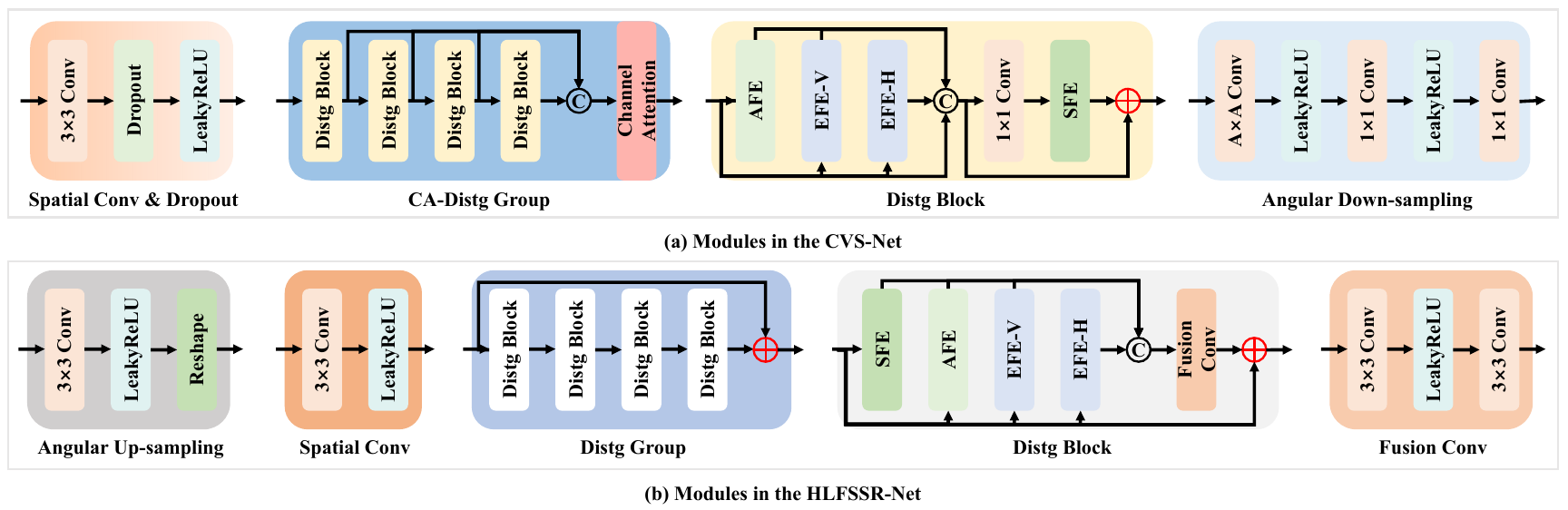}
    \caption{Detailed architectures of the modules in the proposed CVS-Net and HLFSSR-Net. 
    % The Distg Block in (a) is from DistgASR \cite{wang2022disentangling}, while the Distg Block in (b) is from DistgSSR \cite{wang2022disentangling}. The SFE, AFE, EFE-V, and EFE-H represent the spatial, angular, vertical EPI, and horizontal EPI feature extractors. These four feature extractors can disentangle 4D LFs into different 2D sub-spaces, and their cooperation can fully extract the 4D features of LFs.
    }
    \label{fig:archi}
\end{figure*}

As shown in Figure \ref{fig:archi}, we present the detailed architectures of the modules in the proposed CVS-Net and HLFSSR-Net. The Distg Block in Figure \ref{fig:archi}(a) is from DistgASR \cite{wang2022disentangling}, and the Distg Block in Figure \ref{fig:archi}(b) is from DistgSSR \cite{wang2022disentangling}. The SFE, AFE, EFE-V, and EFE-H represent the spatial, angular, vertical EPI, and horizontal EPI feature extractors. These four feature extractors can disentangle 4D LFs into different 2D sub-spaces, and their cooperation can fully extract the 4D features of LFs. In addition, as shown in Table \ref{tab:HLFSSR_archi}, we provide the detailed network architecture of the proposed HLFSSR-Net, including the 2D branch, the 4D branch, and the hybrid LF features extraction and fusion layer. We present the input and output sizes of each layer, and the size of the convolution kernel.
\par
\textbf{Remark}. Our method uses $\mathcal{L}^n$ and ${I}_{\mathbf{u}_0}^h$ as input for training and inference of HLFSSR-Net. Our method only requires the participation of the CVS-Net and BD-Net during the training phase of HLFSSR-Net. In the inference phase, the CVS-Net and BD-Net are no longer required, and the HLFSSR-Net alone can directly perform SR on actual LR LF images $\mathcal{L}^n$.

\begin{table}[htbp]
    \centering
    \renewcommand\arraystretch{1.25}
    \caption{Network architecture of the HLFSSR-Net, where $b$ denotes the batch size, $c$ denotes the channel number, $h \times w$ denotes the spatial resolution of LR SAI, $\alpha h \times \alpha w$ denotes the spatial resolution of HR SAI, and $n (= A \times A)$ denotes the angular number.}
    \label{tab:HLFSSR_archi}
    \resizebox{\linewidth}{!}{
    \centering
    \renewcommand\arraystretch{1.2}
    \begin{tabular}{c|c|c|c|l|c|c|c}
   
    \toprule[1.5pt]
        ~ & \multicolumn{4}{c|}{\textbf{Module}} & \makecell[c]{\textbf{Input size}\\($b\times c\times h\times w$)} & \makecell[c]{\textbf{Kernel size}\\($k\times k\times c_{in} \times c_{out}$)}& \makecell[c]{\textbf{Output size}\\($b\times c\times h\times w$)} \\ 
        \hline\hline
        \multirow{27}{*}{\textbf{2D Branch}} & \multicolumn{3}{c|}{\multirow{2}{*}{Spatial-Conv}} & Conv2d & $b\times 1\times \alpha h\times \alpha w $& $3\times 3\times 1\times n$ & $b\times n\times \alpha h\times \alpha w$ \\ 
        \cline{5-8}
        ~ & \multicolumn{3}{c|}{} & \makecell[l]{LeakyReLU\\(negative slope=0.1)} & - & - & - \\ 
        \cline{2-8}
        ~ & \multicolumn{4}{c|}{LF2MacPI} & $b\times n\times \alpha h\times \alpha w$ & - & $b\times 1\times \alpha A h\times \alpha A w$ \\ 
        \cline{2-8}
        ~ & \multicolumn{4}{c|}{Spatial-Conv}& $b\times 1\times \alpha A h\times \alpha A w$ & $3\times 3\times 1\times 64$ & $b\times 64\times \alpha A h\times \alpha A w$ \\ \cline{2-8}
        ~ & \multirow{23}{*}{Distg-Group$(\times 1)$} & \multirow{20}{*}{Distg-Block} & \multirow{4}{*}{SFE} & Conv2d & $b\times 64\times \alpha A h\times \alpha A w$ & $3\times 3\times 64\times 64$ & $b\times 64\times \alpha A h\times \alpha A w$ \\ 
        \cline{5-8}
        ~ & ~ & ~ & ~ & LeakyReLU (0.1) & - & - & - \\ 
        \cline{5-8}
        ~ & ~ & ~ & ~ & Conv2d & $b\times 64\times \alpha A h\times \alpha A w$ & $3\times 3\times 64\times 64$ & $b\times 64\times \alpha A h\times \alpha A w$ \\ 
        \cline{5-8}
        ~ & ~ & ~ & ~ & LeakyReLU (0.1) & - & - & - \\ 
        \cline{4-8}
        ~ & ~ & ~ & \multirow{5}{*}{AFE} & Conv2d & $b\times 64\times \alpha A h\times \alpha A w$ & $A\times A\times 64\times 16$ & $b\times 16\times \alpha  h\times \alpha  w$ \\ 
        \cline{5-8}
        ~ & ~ & ~ & ~ & LeakyReLU (0.1) & - & - & - \\ 
        \cline{5-8}
        ~ & ~ & ~ & ~ & Conv2d & $b\times 16\times \alpha  h\times \alpha  w$ & $1\times 1\times 16\times 16A^2$ & $b\times 16A^2\times \alpha  h\times \alpha  w$ \\ \cline{5-8}
        ~ & ~ & ~ & ~ & LeakyReLU (0.1) & - & - & - \\ 
        \cline{5-8}
        ~ & ~ & ~ & ~ & PixelShuffle & $b\times 16A^2\times \alpha  h\times \alpha  w$ & - & $b\times 16\times \alpha A h\times \alpha A w$ \\ 
        \cline{4-8}
        ~ & ~ & ~ & \multirow{5}{*}{EFE-H} & Conv2d & $b\times 64\times \alpha A h\times \alpha A w$ & $1\times A^2\times 64\times 32$ & $b\times 32\times \alpha A h\times \alpha w$ \\
        \cline{5-8}
        ~ & ~ & ~ & ~ & LeakyReLU (0.1) & - & - & - \\ 
        \cline{5-8}
        ~ & ~ & ~ & ~ & Conv2d & $b\times 32\times \alpha A h\times \alpha w$ & $1\times 1\times 32\times 32A$ & $b\times 32A\times \alpha A h\times \alpha w$ \\ 
        \cline{5-8}
        ~ & ~ & ~ & ~ & LeakyReLU (0.1) & - & - & - \\ 
        \cline{5-8}
        ~ & ~ & ~ & ~ & PixelShuffle1D & $b\times 32A\times \alpha A h\times \alpha w$ & - & $b\times 32\times \alpha A h\times \alpha A w$ \\ 
        \cline{4-8}
        ~ & ~ & ~ & \multicolumn{2}{c|}{EFE-V}& $b\times 64\times \alpha A h\times \alpha A w$ & - & $b\times 32\times \alpha A h\times \alpha A w$ \\ 
        \cline{4-8}
        ~ & ~ & ~ & \multicolumn{2}{c|}{Cat} & \makecell[c]{$b\times 64\times \alpha A h\times \alpha A w$\\$b\times 16\times \alpha A h\times \alpha A w$\\$b\times 32\times \alpha A h\times \alpha A w$\\$b\times 32\times \alpha A h\times \alpha A w$} & - & $b\times 144\times \alpha A h\times \alpha A w$ \\ 
        \cline{4-8}
        ~ & ~ & ~ & \multirow{3}{*}{Fusion} & Conv2d & $b\times 144\times \alpha A h\times \alpha A w$ & $1\times 1\times 144\times 64$ & $b\times 64\times \alpha A h\times \alpha A w$ \\
        \cline{5-8}
        ~ & ~ & ~ & ~ & LeakyReLU (0.1) & - & - & - \\ 
        \cline{5-8}
        ~ & ~ & ~ & ~ & Conv2d & $b\times 64\times \alpha A h\times \alpha A w$ & $3\times 3\times 64\times 64$ & $b\times 64\times \alpha A h\times \alpha A w$ \\ 
        \cline{3-8}
        ~ & ~ & \multicolumn{3}{c|}{Distg-Block} & $b\times 64\times \alpha A h\times \alpha A w$ & - & $b\times 64\times \alpha A h\times \alpha A w$ \\ 
        \cline{3-8}
        ~ & ~ & \multicolumn{3}{c|}{Distg-Block} & $b\times 64\times \alpha A h\times \alpha A w$ & - & $b\times 64\times \alpha A h\times \alpha A w$ \\ 
        \cline{3-8}
        ~ & ~ & \multicolumn{3}{c|}{Distg-Block} & $b\times 64\times \alpha A h\times \alpha A w$ & - & $b\times 64\times \alpha A h\times \alpha A w$ \\ 
        \cline{2-8}
        ~ & \multicolumn{4}{c|}{Spatial-Conv} & $b\times 64\times \alpha A h\times \alpha A w$ & $3\times 3\times 64\times 1$ & $b\times 1\times \alpha A h\times \alpha A w$ \\ 
        \hline\hline
        \multirow{5}{*}{\textbf{4D Branch}} & \multicolumn{4}{c|}{Spatial-Conv} & $b\times 1\times Ah\times Aw$ & $3\times 3\times 1\times 64$ & $b\times 64\times Ah\times Aw$ \\ 
        \cline{2-8}
        ~ & \multicolumn{4}{c|}{Distg-Group $(\times 4)$} & $b\times 64\times Ah\times Aw$ & - & $b\times 64\times Ah\times Aw$ \\ 
        \cline{2-8}
        ~ & \multicolumn{3}{c|}{\multirow{3}{*}{Up-sampling}} & Conv2d & $b\times 64\times Ah\times Aw$ & $1\times 1\times 64\times 256$ & $b\times 256 \times Ah\times Aw$ \\ 
        \cline{5-8}
        ~ & \multicolumn{3}{c|}{} & PixelShuffle & $b\times 256 \times Ah\times Aw$ & - & $b\times 64\times \alpha A h\times \alpha A w$ \\ 
        \cline{5-8}
        ~ & \multicolumn{3}{c|}{} & Conv2d & $b\times 64\times \alpha A h\times \alpha A w$ & $1\times 1\times 64\times 1$ & $b\times 1\times \alpha A h\times \alpha A w$ \\ 
        \hline\hline
        \multirow{6}{*}{\textbf{Hybrid Features Fusion}} & \multicolumn{4}{c|}{Cat} & \makecell[c]{$b\times 1\times \alpha A h\times \alpha A w$ \\ $b\times 1\times \alpha A h\times \alpha A w$} & - & $b\times \alpha \times \alpha A h\times \alpha A w$ \\ 
        \cline{2-8}
        ~ & \multicolumn{4}{c|}{Spatial-Conv} & $b\times 2\times \alpha A h\times \alpha A w$ & $3\times 3\times 2\times 64$ & $b\times 64\times \alpha A h\times \alpha A w$ \\ 
        \cline{2-8}
        ~ & \multicolumn{4}{c|}{Distg-Group $(\times 1)$} & $b\times 64\times \alpha A h\times \alpha A w$ & - & $b\times 64\times \alpha A h\times \alpha A w$ \\ 
        \cline{2-8}
        ~ & \multicolumn{3}{c|}{\multirow{3}{*}{Fusion-Conv}} & Conv2d & $b\times 64\times \alpha A h\times \alpha A w$ & $3\times 3\times 64\times 16$ & $b\times 16\times \alpha A h\times \alpha A w$ \\ 
        \cline{5-8}
        ~ & \multicolumn{3}{c|}{} & LeakyReLU (0.1) & - & - & - \\ 
        \cline{5-8}
        ~ & \multicolumn{3}{c|}{} & Conv2d & $b\times 16\times \alpha A h\times \alpha A w$ & $3\times 3\times 16\times 1$ & $b\times 1\times \alpha A h\times \alpha A w$ \\ 
        \bottomrule[1.5pt]
    \end{tabular}
    
    }
    
\end{table}

\section{lf reorganization}
\label{sec:s2}
\begin{figure}[htbp]
    \centering
    \includegraphics[width=\linewidth]{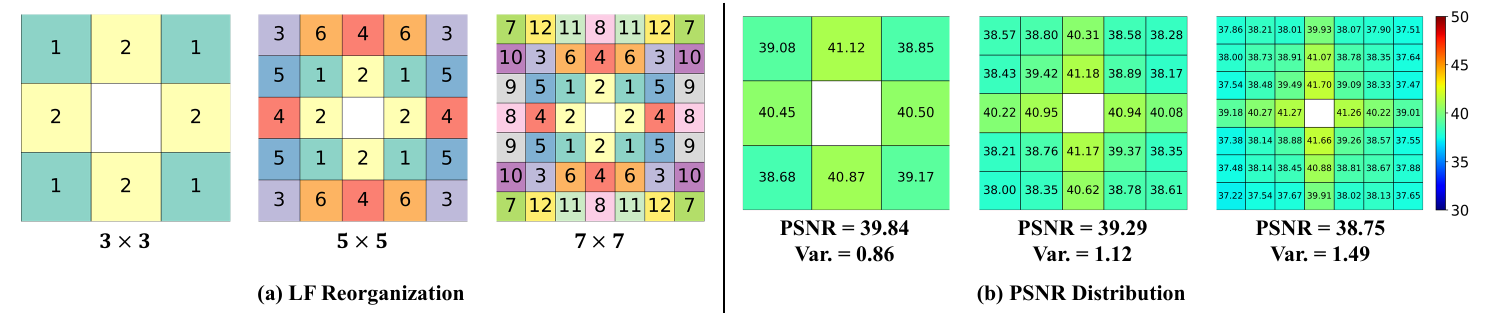}
    \caption{The LF reorganization (a), and the average PSNR distribution (b) on the HCI new test set among different side-view SAIs under different angular resolutions.}
    \label{fig:compareDiffAngPSNR}
\end{figure}

As shown in Figure \ref{fig:compareDiffAngPSNR}(a), we list the LF reorganization strategies with different angular resolutions. The views indicated by the same numbers in the Figure \ref{fig:compareDiffAngPSNR}(a) will be assigned to the same set. As shown in Figure \ref{fig:compareDiffAngPSNR}(b), we provide the average PSNR distribution among different side-view SAIs on the HCI new dataset obtained by our method for $2\times$ SR. 
% It can be seen that the views closer to the center have higher PSNR scores. 
Moreover, we evaluate the performance of our SR method with different angular resolutions. Specifically, we extract the central $A\times A$ ($A = 3,5,7$) SAIs from the input LFs and train corresponding models for $2 \times$ and $4 \times$ SR. As shown in Table \ref{tab:Ab4AngRes}, the PSNR values tend to decrease in a small range when the angular resolution increases. That is because there is no HR ground truth for the side-view SAIs, and the quality of super-resolved side-view SAIs is directly related to the performance of the CVS-Net. The central view synthesized from the views far from the center is not as easy as that synthesized from the views closer to the center. As shown in Figure \ref{fig:compareDiffAngPSNR}, the farther the side-view SAIs are from the center, the lower their PSNR scores are, these results are foreseeable and acceptable. In future work, we will further improve the SR results of our method by refining the CVS-Net.

\begin{table}[htbp]
\centering
\renewcommand\arraystretch{1.25}
    \caption{The parameters, average PSNR/SSIM scores and the variances in different view's PSNR achieved by our proposed method with different angular resolutions for $2\times$ and $4\times$ SR.}
    \label{tab:Ab4AngRes}
    %\vspace{-0.3cm}
\resizebox{0.5\linewidth}{!}
{
    \centering
    \begin{tabular}{c|c|c|c|c|c|c}
    \toprule[1.2pt]      
         \multirow{2}*{Scale} & \multirow{2}*{AngRes} & \multirow{2}*{\#Param.} & \multicolumn{2}{c|}{HCI new} & \multicolumn{2}{c}{DLFD} \\
         \cline{4-7}
         ~ & ~ & ~ & PSNR/SSIM & Var. & PSNR/SSIM & Var. \\
         \hline
         \multirow{3}*{$2\times$} & $3\times3$ & 5.98M  & 39.84/0.985 & 0.86 & 42.01/0.989 & 0.78\\
         ~ & $5\times5$ & 7.97M  & 39.29/0.984 & 1.12 & 41.30/0.987 & 1.20 \\
         ~ & $7\times7$ & 10.92M  & 38.75/0.984 & 1.49 & 40.57/0.987 & 1.87
 \\     
         \hline
         \multirow{3}*{$4\times$} & $3\times3$ & 5.99M  & 37.08/0.978  & 1.67 & 38.01/0.978 & 3.27 \\
         ~ & $5\times5$ & 7.98M  & 35.30/0.969 & 4.42 & 36.20/0.966 & 4.94 \\
         ~ & $7\times7$ & 10.93M  & 32.43/0.946 & 6.53 & 33.33/0.943 & 8.53 \\     
         \bottomrule[1.2pt]
    \end{tabular}
    }
\end{table}
\par

\par

\section{ablation study on EPI gradient loss}
\label{sec:s3}
To further validate the contribution of the epipolar-plane image (EPI) gradient loss \cite{jin2020AlltoOne,jin2020LFASR} to maintaining geometric consistency, we conducted an additional ablation experiment based on the results reported in Table \textcolor{red}{\Rmnum{3}} in the manuscript. Specifically, we replaced the EPI gradient loss between $\hat{\mathcal{L}}^n$ and $\mathcal{L}^n$ in Figure \textcolor{red}{4} with a standard L1 loss, and retrained the HLFSSR-Net on the simulated datasets (HCI new \cite{honauer2017HCI} and DLFD \cite{shi2019DLFD}). The comparison with the complete framework (Framework 4 in Table \textcolor{red}{\Rmnum{3}}) is summarized in Table \ref{tab:ablation_EPI}.
\par
The results show that the utilization of the EPI gradient loss (w/ EPI gradient loss) leads to an improvement of approximately 0.3--0.4 dB on both the HCI new and DLFD datasets, compared with using the L1 loss (w/o EPI gradient loss). This clearly demonstrates the effectiveness of the EPI gradient loss in preserving LF parallax structures and maintaining geometric consistency.

\begin{table}[ht]
\centering
\renewcommand\arraystretch{1.25}
    \caption{Supplementary ablation study on EPI gradient loss. The PSNR/SSIM scores of $2\times$ SR results on $5 \times 5$ LFs are provided. }
    \label{tab:ablation_EPI}
    %\vspace{-0.3cm}
{
    \centering
    \begin{tabular}{c|c|c}
    \toprule[1.2pt]
         ~ & w/o EPI gradient loss & w/ EPI gradient loss  \\
         \hline
         HCI new \cite{honauer2017HCI}  & 38.91/0.982 &  39.29/0/984  \\
         DLFD \cite{shi2019DLFD} &  40.85/0.985   & 41.30/0.987   \\
         \bottomrule[1.2pt]
    \end{tabular}
    }
\end{table}

\section{sr performance on the proposed real-world hybrid LF dataset}
\label{sec:s4}

\begin{figure*}[htbp]
    \centering
    \includegraphics[width=0.9\linewidth]{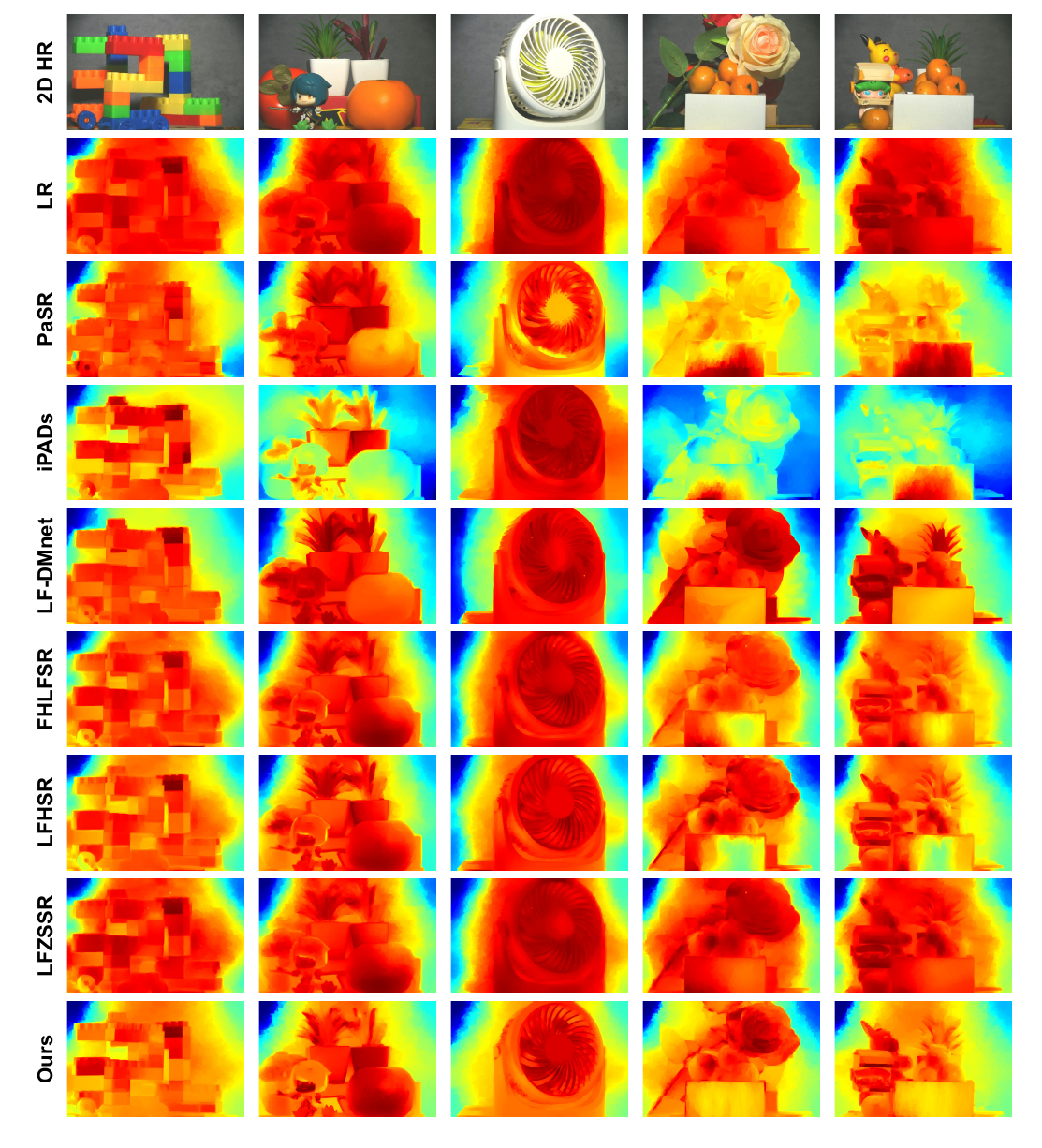}
    \caption{Visual comparisons of the estimated disparity maps from the LR LF images and super-resolved LF images by different methods on our real-world hybrid data.}
    \label{fig:compareDepth}
\end{figure*}
We visually compare the disparity maps, as illustrated in Figure \ref{fig:compareDepth}, which are estimated from the LR LF images and the $2\times$SR LF images by different methods using an identical LF disparity estimation algorithm~\cite{chen2018depth} with angular resolution of $5\times5$. The better disparity maps in Figure \ref{fig:compareDepth} illustrate the ability of our SR method to improve spatial resolution while preserving the inherent LF parallax structure simultaneously. 
\par
It is worth noting, as shown in Table \ref{tab:dataset}, the number and type of scenes in our real-word hybrid LF dataset was determined with reference to the publicly available datasets commonly used for training and evaluating state-of-the-art LF spatial SR algorithms \cite{wang2022disentangling, jin2023LFHSR}. These datasets were also employed in our work to construct the simulated hybrid LF dataset described in the manuscript. Accordingly, the scene configurations and scale of our dataset were designed to be consistent with these benchmarks. However, the current dataset, which includes 189 real-world indoor scenes, may impose certain limitations when generalizing to outdoor scenes with large depth variations. A more extensive dataset encompassing a wider variety of scenes would further enhance the generalization ability of the proposed method. To address this, we have initiated work on extending our dataset by (1) providing accurate depth ground truth for similar LF scenes, and (2) improving the portability of our hybrid LF imaging prototype to facilitate the acquisition of a richer set of indoor and outdoor scenes. Constructing this expanded dataset is an important part of our future work.
\begin{table}[h]
\centering
\renewcommand\arraystretch{1.25}
    \caption{The composition of the simulated datasets.}
    \label{tab:dataset}
    %\vspace{-0.3cm}
% \resizebox{0.75\linewidth}{!}
{
    \centering
    \begin{tabular}{c|c|c|c}
    \toprule[1.2pt]
         Dataset & Training & Test & Category  \\
         \hline
         HCI new \cite{honauer2017HCI} & 20  & 4  & Synthetic\\
         DLFD \cite{shi2019DLFD} & 20  & 19 & Synthetic \\
         \hline
         EPFL \cite{rerabek2016EPFL}& 70  & 10 & Real-world \\       
         INRIA \cite{le2018INRIA}& 35  & 5 & Real-world \\   
         STFgantry \cite{vaish2008STFgantry}& 9  & 2 & Real-world \\   
         \bottomrule[1.2pt]
    \end{tabular}
    }
\end{table}

\par
\section{quantitative results on domain gap}
\label{sec:s5}
In addition, we evaluate the domain gap quantitatively with the PSNR of each method on the training set and the test set. For example, as shown in Table \ref{tab:domaingap}, in the $2 \times$ SR experiments on the HCI new dataset, the average PSNR of LFHSR on the training set converges to nearly 42 dB, and the average PSNR of LFHSR \cite{jin2023LFHSR} on the test set is 40.28 dB as shown in Table \ref{tab:domaingap}, with the difference in PSNR between the training and test stages is 1.63 dB. In our method, the PSNR on the training set and test set are 40.84dB and 39.29dB, respectively, with a difference of 1.55 dB. However, in the $4 \times$ SR experiment, the average PSNR of \textbf{LFHSR} \cite{jin2023LFHSR} converges to nearly 39 dB on the training set, but its average PSNR on the test set is only 31.84 dB, with a \textbf{PSNR difference of 6.92 dB} between the training and test sets. In contrast, the average PSNR of \textbf{our method} on the training and test sets is 36.50 dB and 35.30 dB, respectively, with a difference of \textbf{only 1.20 dB}. The PSNR difference of LFHSR \cite{jin2023LFHSR} in the $4 \times$ SR task is much larger than that in the $2 \times$ SR task, whereas our method's PSNR difference in both the $2 \times$ and the $4 \times$ SR tasks is within normal limits. This result can be used as a quantitative basis for further explaining the advantages of the self-supervised method under specific degradation conditions.
\begin{table}[h]
    \centering
    \caption{Differences ($\Delta$PSNR) between training set PSNR and test set PSNR.}
    % \resizebox{\linewidth}{!}
    {
    \begin{tabular}{c|ccc|ccc}
    \toprule[1.2pt]
         \multirow{2}*{}  & \multicolumn{3}{c|}{$2 \times$ SR} & \multicolumn{3}{c}{$4 \times$ SR}  \\
         \cline{2-7}
         ~ & Training & Test &	$\Delta$PSNR & Training & Test &	$\Delta$PSNR \\
         \hline
         LFHSR \cite{jin2023LFHSR} & 41.91 dB &	40.28 dB & 1.63 dB & 38.76 dB & 31.84 dB	& \textcolor{red}{6.92} dB \\
         \hline
         Ours & 40.84 dB	& 39.29 dB	& 1.55 dB & 36.50 dB	& 35.30 dB &	\textbf{1.20} dB \\
         \bottomrule[1.2pt]
         
    \end{tabular}
    }
    \label{tab:domaingap}
\end{table}

% and the distribution of the scores is relatively uniform. This further points the way for our future work to refine the CVS-Net.

\par
% \noindent{\textbf{REFERENCE}}
% \par
% \noindent[\textcolor{green}{A}] Wang \textit{et al.}, Disentangling light fields for super-resolution and disparity estimation, TPAMI 2022.

% \begin{table}[!]
% \centering
% \renewcommand\arraystretch{1.25}
%     \caption{The composition of the simulated datasets.}
%     \label{tab:dataset}
%     %\vspace{-0.3cm}
% % \resizebox{0.75\linewidth}{!}
% {
%     \centering
%     \begin{tabular}{c|c|c|c}
%     \toprule[1.2pt]
%          Dataset & Training & Test & Category  \\
%          \hline
%          HCI new \cite{honauer2017HCI} & 20  & 4  & Synthetic\\
%          DLFD \cite{shi2019DLFD} & 20  & 19 & Synthetic \\
%          \hline
%          EPFL \cite{rerabek2016EPFL} & 70  & 10 & Real-world \\       
%          INRIA \cite{le2018INRIA} & 35  & 5 & Real-world \\   
%          STFgantry \cite{vaish2008STFgantry} & 9  & 2 & Real-world \\   
%          \bottomrule[1.2pt]
%     \end{tabular}
%     }
% \end{table}

% \fi
\IEEEpeerreviewmaketitle

%%=============================================================%%
%%======================  Introduction   ======================%%
%%=============================================================%%

% or


% \appendices

% \section{}

% \section*{Acknowledgment}


\ifCLASSOPTIONcaptionsoff
  \newpage
\fi

\normalem

\bibliographystyle{IEEEtran}
\bibliography{bibtex/bib/IEEEabrv,bibtex/bib/bib}